\documentclass[a4paper,11pt]{article}
\pdfoutput=1

\usepackage{a4wide} 
\usepackage{rotating}

\usepackage[latin1]{inputenc}
\usepackage{amsfonts}
\usepackage{amsmath}
\usepackage{amssymb}
\usepackage{amsthm}
\usepackage{graphicx}
\usepackage{booktabs}
\usepackage{accents}
\usepackage{color}
\usepackage{cite}
\usepackage[small,bf]{caption}
\usepackage[small]{subfigure}

\newcommand{\ord}[1]{\mathcal{O}\left(#1 \right)}

\allowdisplaybreaks[1]

\begin{document}

\begin{titlepage}

\vspace*{-15mm}
\begin{flushright}
MPP-2012-120\\
SISSA 22/2012/EP
\end{flushright}
\vspace*{0.7cm}

\begin{center}
{
\bf\LARGE
Naturalness of the Non-Universal MSSM \\[0.5ex]
in the light of the recent Higgs results}
\\[8mm]
Stefan~Antusch$^{\star, \dagger}$
\footnote{E-mail: \texttt{stefan.antusch@unibas.ch}},
Lorenzo~Calibbi$^{\dagger}$
\footnote{E-mail: \texttt{calibbi@mppmu.mpg.de}},
Vinzenz~Maurer$^{\star}$
\footnote{E-mail: \texttt{vinzenz.maurer@unibas.ch}},\\
Maurizio~Monaco$^{\ddag}$
\footnote{E-mail: \texttt{mmonaco@sissa.it}},
Martin~Spinrath$^{\ddag}$
\footnote{E-mail: \texttt{spinrath@sissa.it}}
\\[1mm]
\end{center}
\vspace*{0.50cm}
\centerline{$^{\star}$ \it
 Department of Physics, University of Basel,}
\centerline{\it
Klingelbergstr.~82, CH-4056 Basel, Switzerland}
\vspace*{0.2cm}
\centerline{$^{\dagger}$ \it
Max-Planck-Institut f\"ur Physik (Werner-Heisenberg-Institut),}
\centerline{\it
F\"ohringer Ring 6, D-80805 M\"unchen, Germany}
\vspace*{0.2cm}
\centerline{$^{\ddag}$ \it
SISSA/ISAS and INFN,}
\centerline{\it
Via Bonomea 265, I-34136 Trieste, Italy }
\vspace*{1.20cm}
\begin{abstract}

\noindent
We analyse the naturalness of the Minimal Supersymmetric Standard Model (MSSM) in the light of recent LHC results from the ATLAS and CMS experiments. We study non-universal boundary conditions for the scalar and the gaugino sector, with fixed relations between some of the soft breaking parameters, and find a significant reduction of fine-tuning for non-universal gaugino masses. For a Higgs mass of about 125~GeV, as observed recently, we find parameter regions with a fine-tuning of $\ord{10}$, taking into account experimental and theoretical uncertainties. These regions also survive after comparison with simplified model searches in ATLAS and CMS. For a fine-tuning less than 20 the lightest neutralino is expected to be lighter than about 400 GeV and the lighter stop can be as heavy as 3.5 TeV. On the other hand, the gluino mass is required to be above 1.5 TeV. For non-universal gaugino masses, we discuss which fixed GUT scale ratios can lead to a reduced fine-tuning and find that the recent Higgs results have a strong impact on which ratio is favoured. We also discuss the naturalness of GUT scale Yukawa relations, comparing the non-universal MSSM with the CMSSM.  
\end{abstract}

\end{titlepage}

\section{Introduction}
Recently, the LHC experiments ATLAS and CMS have reported the discovery of a new resonance, mainly based on excesses of events in the $\gamma\gamma$ and $ZZ^*$ channels. The combined significance amounts about $5 \sigma$ for both, ATLAS and CMS \cite{discovery}. The new particle is compatible with the Standard Model (SM) Higgs boson with a mass of $m_h\approx 125 \div 126$~GeV~\cite{Higgs-fits}.
In the context of low-energy Supersymmetry (SUSY), such a value still lies in the predicted range of the Minimal Supersymmetric Standard Model (MSSM), provided quite heavy top squarks (with masses $\gtrsim \ord{1}$ TeV) and/or large left-right stop mixing (see for instance \cite{Arbey:2011ab}) are present. 
Regarding direct searches for SUSY particles \cite{SMSbounds}, based on $\sim 5$ fb$^{-1}$ of data, only negative results have been reported. 

From the point of view of the MSSM, the negative results of direct SUSY searches and the comparatively heavy Higgs mass are well consistent with each other. However, as the main motivation for SUSY is the solution to the hierarchy problem, which naively requires SUSY masses of the order of the electroweak symmetry breaking scale, one may ask whether the MSSM can still accomplish this task in a natural way (or at least for a moderate fine-tuning price). 
Regarding specific MSSM scenarios, it turns out that in the constrained MSSM (CMSSM) \cite{Chamseddine:1982jx} a significant amount of fine-tuning (at least $\gtrsim {\cal O}(100)$) is unavoidable to explain a Higgs mass of $m_h\approx 125 \div 126$~GeV. Following our paper \cite{Antusch:2011xz} on the CMSSM, 
we thus find it interesting to re-address the naturalness of the MSSM in the context of models where certain universality assumptions on the SUSY parameters are relaxed 
(e.g.~for scalar and gaugino masses). 

In order to envisage the most promising ways to reduce the fine-tuning, we are going to adopt the following strategy: 
we start with considering a generic setup with 17 independent parameters inspired by the phenomenological MSSM (pMSSM), in order to identify those giving the dominant contributions to the fine-tuning.
This allows us to identify the rigid relations among the parameters which can decrease fine-tuning. 
After discussing several possibilities, we will mainly focus on certain non-universal relations among gaugino masses, whose possible impact on fine-tuning have been 
studied before the LHC results by \cite{Abe:2007kf, Gogoladze:2009bd, Horton:2009ed}.\footnote{A recent discussion of the fine-tuning price of a 125 GeV Higgs after the first hints in December 2011 within several SUSY models has been also given in \cite{Ghilencea:2012gz} (see also \cite{Brummer:2012zc, Hall:2011aa}).}
In this context, we will study the interplay between the fine-tuning, the model predictions for the light Higgs mass and the Grand Unified Theory (GUT) scale ratios of the gaugino masses and of the third family Yukawa couplings, which are key quantities for discriminating among different SUSY (and SUSY GUT) models. In our analysis, we include various relevant experimental constraints, e.g.\ from BR$(b \to s \gamma)$, BR$(B_s \to \mu^+ \mu^-)$ and BR$(B_u \to \tau \nu_\tau)$. We do not impose strict constraints from requiring that the $(g-2)_\mu$ deviation from the SM (currently at the level of $3.2 \sigma$ \cite{g-2}) is explained by SUSY. 

The rest of the paper is organized as follows: in the next section we study
in a semi-analytic way the dependence of the electroweak scale on GUT scale
boundary conditions whose choice is inspired by the pMSSM.
Thanks to this analysis we can identify regions with low fine-tuning for certain
limiting cases and identify as most promising case non-universal gaugino masses.
In section 3 we revisit briefly the dependence of GUT scale Yukawa coupling
ratios on low energy supersymmetric threshold corrections. Thereafter
we make an extensive numerical analysis of the fine-tuning in an MSSM scenario
with non-universal gaugino masses at the GUT scale and compare the results
to various experimental results, most importantly the recent discovery of a
new resonance around 125~GeV compatible with a Higgs boson. In the final
section 5 we summarize and conclude.

\section{Fine-Tuning in the MSSM} \label{sec:FineTuning}

Fine-tuning in the MSSM has been extensively discussed
in the literature starting with \cite{Barbieri:1987fn}.\footnote{For an
extensive list of references, see for instance \cite{arXiv:1110.6926, Ghilencea:2012gz}.}
As we are going to follow the same approach as in our recent paper \cite{Antusch:2011xz},
we give here just a brief summary of the most important points.

In the MSSM, by minimizing the scalar potential,  
the $Z$-boson mass can be computed in terms of the
supersymmetric Higgs parameter $\mu$ and the soft SUSY breaking mass terms
of the up- and down-type Higgs doublets.
For moderate and large $\tan \beta$, one finds at tree level 
\begin{equation}
    \frac{M_Z^2}{2} = -|\mu|^2 - m^2_{H_u} + \mathcal{O}(m^2_{H_{u,d}} / (\tan\beta)^2) \;. \label{eq:Zbosonmass_expansion}
\end{equation}
The value of the low energy observable $M_Z$ can thus be
obtained in terms of the fundamental parameters of the high-energy theory, 
by considering the renormalization group (RG) evolution of $\mu$
and $m_{H_u}$, which are assumed to be both of the order of
the SUSY breaking mass scale, from the GUT scale down to low energy. 
Since experimental constraints 
force the SUSY scale to be somehow larger than $M_Z$, a certain amount of 
tuning is needed.
In order to quantify fine-tuning, the following measure has been
introduced \cite{Barbieri:1987fn}
\begin{equation}\label{eq:BarbieriFTdefinition}
    \Delta_a = \left| \frac{\partial \log M_Z}{\partial \log a} \right| = \left|\frac{a}{2 M_Z^2} \frac{\partial M_Z^2}{\partial a} \right| \;.
\end{equation}
$\Delta_a$ reflects the dependence of $M_Z$ on the variation of a
given GUT-scale Lagrangian parameter $a$. This definition not only
encompasses the obvious fine-tuning in $\mu$ which is needed 
to fulfill Eq.~\eqref{eq:Zbosonmass_expansion}, but it also
covers the tuning needed to have a small $m^2_{H_u}$, while
other soft SUSY breaking parameters entering its RG evolution are
relatively large.
The overall measure of fine-tuning for a given parameter point
is then defined as the maximum of all the single $\Delta_a$'s: 
\begin{equation}
    \Delta = \max_{\substack{a}} \Delta_a \;. \label{eq:wholetuning}
\end{equation}
Turning to the contributions $\Delta_a$ of the single parameters, 
let us first notice that the fine-tuning in $\mu$ is rather special 
as its low-energy value is determined by imposing the correct $M_Z$. 
Hence, setting aside RG effects for the moment, $\Delta_\mu$ can be approximately expressed in terms of $m^2_{H_u}$ as 
\begin{equation} \label{eq:FTmu}
    \Delta_\mu \approx \left| 2 \frac{m^2_{H_u}}{M_Z^2} + 1 \right| \,.
\end{equation}
In the numerical analysis, the running of $\mu$ is of course included. 
It turns out that in the constrained MSSM $\Delta_\mu$ is often the
dominant term in $\Delta$ followed by the tuning in $A_0$, $m_0$ and $M_{1/2}$.

The discussion of fine-tuning for the other parameters is more involved 
as they enter Eq.\ \eqref{eq:Zbosonmass_expansion} 
only via the RG evolution of $m^2_{H_{u,d}}$. It is, however, possible to 
approximately express $m^2_{H_u}$ as a polynomial in terms of the
high-energy parameters of the theory.

The coefficients of this polynomial depend strongly on the top Yukawa coupling
and the strong gauge coupling constant.  Therefore it was argued in the literature
that these two parameters should be included in the fine-tuning measure, see, e.g.\
\cite{hep-ph/9303291}. But since they are measured, in contrast to the
supersymmetric parameters, there exist different approaches how to implement them
into the fine-tuning measure. We will follow here the approach to weight them with
their experimental uncertainty $\sigma$
\begin{equation}
 \Delta_{y_t} = \left|\frac{\sigma_{y_t}}{2 M_Z^2} \frac{\partial M_Z^2}{\partial y_t} \right| \;,
\end{equation}
and similarly for the strong coupling constant. Note that we have to take the evolved 
experimental error at the GUT scale, including RG running, which is bigger than the lower
energy one for the top Yukawa coupling and smaller for the strong coupling
constant. In our numerical analysis later on we use the mean
values at low energies $m_t = 172.9 \pm 0.6 \pm 0.9$~GeV (the top quark pole mass) \cite{PDG}
and $\alpha_s (M_Z) = 0.1184 \pm 0.0007$ \cite{PDG} and the relative uncertainties
$\sigma_{y_t} = 4\%$ and $\sigma_{\alpha_s} = 0.2\%$ at the GUT scale.

In the following we will give approximate
formulas for the dependence of $m^2_{H_u}$ on the assumed fundamental
soft SUSY breaking GUT-scale parameters at low energies, which can tell us
for which choice of parameters we can expect a reduced fine-tuning. 
We remark that we will not include the dependence on $y_t$ and $\alpha_s$ in 
the analytical discussion, since this would make the formulas too involved. 
An analytic discussion of fine-tuning in $y_t$ and $\alpha_s$ is therefore beyond 
the scope here, but the effect will be included in the numerical analysis as described above. 

Before we come to the analytical discussion, we also like to note that fine-tuning measures
should be taken with some caution.
It is a matter of individual taste how much fine-tuning one accepts as ``natural'' and it also depends on 
the fine-tuning definition used. 
Thus we want to stress that our aim here is mainly to compare different parameter regions and only 
the relative fine-tuning difference between them would render a region more attractive to us.

\subsection{Fine-Tuning \`a la pMSSM}

To express $m^2_{H_u}$ in terms of high-energy parameters, our  
choice of the independent parameters is
inspired by the so-called phenomenological MSSM (pMSSM) \cite{Djouadi:1998di}
(but with the parameters defined at the GUT scale).
In particular, we consider different scalar masses for the chiral superfields
of the MSSM. The first two generation sfermions are assumed to be degenerate,
while we allow a splitting for the third generation. The gaugino masses are
taken to be non-universal. Concerning the trilinears we introduce three independent
parameters: one for the up squarks, one for the down squarks and one for
the charged sleptons. The main difference to the usual pMSSM is in the Higgs sector. 
Namely, instead of low scale $\mu$ and the CP-odd Higgs mass $m_{A_0}$, we
take the GUT scale value of $m^2_{H_u}$ and $m^2_{H_d}$ to be the free parameters
and denote them by $m^2_{h_u}$ and $m^2_{h_d}$ respectively.

The dependence of $m^2_{H_u}$ on the GUT scale parameters can be deduced from its RGE 
(see e.g.\ \cite{Martin:1993zk}) and can be written as
 \begin{equation}
 m^2_{H_u} = \sum_i a_i m_i^2 + \frac{1}{2} \sum_{i,j} N_i b_{ij} N_j \;,
\label{eq:param}
 \end{equation}
where the $N_i \equiv (M_1, M_2, M_3, A_t, A_b, A_\tau)$ are
assumed to be real and the matrix $b_{ij}$
is a general symmetric 6$\times$6 matrix.

To make the deviation from the universality assumption of the CMSSM
more evident we introduce the following dimensionless
quantities: 
\begin{equation}
 \eta_\alpha = M_\alpha/M_3 \quad \text{with}\,\, \alpha = 1,2,3 \qquad
\qquad \eta_i = A_{i}/M_3 \quad \text{with}\,\, i = t,b,\tau. 
\end{equation}
Note that the convention we use here and in the rest of the paper implies $\eta_3 = 1$ and we can recast 
the $N_i$ as 
\begin{equation}
{N}_i = M_3 \cdot (\eta_1, \eta_2, \eta_3, \eta_t, \eta_b, \eta_\tau)\:.
\end{equation} 
The fine-tuning measure introduced in 
Eq.~\eqref{eq:wholetuning} can now be easily expressed as
\begin{equation}
\label{eq:FT_mi_pMSSM}
\Delta_{m_i^2} = \left| 2 a_i \frac{m_i^2}{M_Z^2} \right|, \qquad \Delta_m = \max_{\substack{i}} \Delta_{m_i^2} \,,
\end{equation} 
for the scalar masses, whereas for the other soft terms we find
\begin{equation}
\label{eq:FT_Ni_pMSSM}
\Delta_{N_i} = \left| \frac{\sum_j {N}_i b_{ij} {N}_j }{M_Z^2} \right| \quad \text{(no sum over $i$)}, \qquad \Delta_N = \max_{\substack{i}} \Delta_{N_i} \,,
\end{equation} 
and
\begin{equation}
\label{eq:FT_mu_pMSSM}
\Delta_\mu \approx \left| 2 \frac{\sum_i a_i m_i^2 + \frac{1}{2} \sum_{i,j} N_i b_{ij} N_j}{M_Z^2} + 1 \right| \,.
\end{equation}

After this general introduction to fine-tuning \`a la pMSSM, we turn to
expressing $m^2_{H_u}$ at the SUSY scale for a point which corresponds
to $M_3= 0.6$~TeV, $m_i = 1.5$~TeV for every $i$, $\eta_i = {N}_i /M_3 = (1,1,1,-5,-5,-5)$, 
$\tan \beta = 30$ and $\mu$ positive. This point is in agreement with the Higgs mass range
quoted \cite{discovery} and the $\text{BR}(B_s \to \mu^+ \mu^-)$ bound \cite{Aaij:2012ac} and
we find using {\tt softSUSY 3.2.3}  \cite{Allanach:2001kg}
\begin{align} \label{eq:mHuSq}
 m^2_{H_u} (M_{\text{SUSY}}) &=
-0.0459 m_{\tilde{Q}_1}^2 + 0.0988 m_{\tilde{U}_1}^2 - 0.0469 m_{\tilde{D}_1}^2 + 0.0488 m_{\tilde{L}_1}^2 - 0.0541 m_{\tilde{E}_1}^2  \nonumber \\
&\quad -0.3347 m_{\tilde{Q}_3}^2 - 0.2500 m_{\tilde{U}_3}^2 - 0.0154 m_{\tilde{D}_3}^2 + 0.0245 m_{\tilde{L}_3}^2 - 0.0236 m_{\tilde{E}_3}^2  \nonumber \\
&\quad + 0.6481 m_{h_u}^2 + 0.0273 m_{h_d}^2 \nonumber \\
&\quad - M_3^2 ( 1.2865 - 0.0216 \eta_1 - 0.0242 \eta_1^2 + 0.0230 \eta_2 - 
 0.2177 \eta_2^2 + 0.0813 \eta_1 \eta_2 ) \nonumber\\
&\quad + M_3^2 (0.2521 \eta_t + 0.0208 \eta_b +  0.0175 \eta_\tau) \nonumber\\
&\quad + M_3^2 \eta_1 (0.0087 \eta_t - 0.0028 \eta_b + 0.0053 \eta_\tau ) \nonumber\\
&\quad + M_3^2 \eta_2 (0.0852 \eta_t - 0.0086 \eta_b - 0.0114 \eta_\tau) \nonumber\\
&\quad + M_3^2 (0.0022 \eta_b^2 - 0.1244 \eta_t^2 - 0.0001 \eta_\tau^2) \nonumber\\
&\quad + M_3^2 (0.0068 \eta_b \eta_t + 0.0017 \eta_b \eta_\tau + 0.0007 \eta_t \eta_\tau) \;,
\end{align}
where $\eta_3$ is set to be 1.
One can already spot here the most important contributions to come from the stops, the
gluinos and the Higgs sector itself.

The coefficients $a_i$ appearing in Eq.~\eqref{eq:param} 
can be directly read off from Eq.~\eqref{eq:mHuSq}. 
In order to show the correlations among the parameters $\eta_i = {N}_i/M_3$ in a transparent way, 
we display the coefficients $b_{ij}$ as a symmetric matrix:
\begin{equation}
 b_{ij} = \begin{pmatrix}
           0.0242 & -0.0813 &  0.0216 &  0.0087 & -0.0028 &  0.0053 \\
	          &  0.2177 & -0.0230 &  0.0852 & -0.0086 & -0.0114 \\
	          &         & -1.2865 &  0.2521 &  0.0208 &  0.0175 \\
	          &         &         & -0.1244 &  0.0068 &  0.0007 \\
	          &         &         &         &  0.0022 &  0.0017 \\
	          &         &         &         &         & -0.0001 
          \end{pmatrix} \;.
\end{equation}
Even though the coefficients $a_i$ and $b_{ij}$ were numerically obtained 
in a specific point of the parameter space, we checked that they 
provide a reasonably accurate estimate of $m^2_{H_u}$ in wide regions of parameter space.
The corresponding uncertainty of the coefficients is estimated to be of the
order of 10-20\%. 
This is good enough for the qualitative discussion we present in the following subsections on possible strategies to reduce the fine-tuning.
Nevertheless, the results we present in section \ref{sec:num} are based on a full numerical analysis, with
the RGE evolution of all parameters computed in each point of the parameter space.

\subsection{Our Strategy}
\label{sec:strategy}

In the attempt to find SUSY models with reduced fine-tuning
we assume that the underlying model of SUSY breaking predicts certain
fixed relations between the SUSY breaking parameters
at high energy (e.g.\ at the GUT scale). As a consequence, cancellations among different 
contributions in Eq.~\eqref{eq:mHuSq} can occur which lead to a reduced fine-tuning. 
We note that such a behaviour is a well-known property of models with universal 
scalar masses (like the CMSSM), for which the coefficients $a_i$ in Eq.~\eqref{eq:mHuSq} 
cancel almost completely.\footnote{This property is known under the name of ``focus point'' \cite{focuspoint}, 
for a recent quantitative discussion see, e.g.\ \cite{Antusch:2011xz}.}

The coefficients of the GUT scale parameters in this equation depend
also strongly on the top Yukawa coupling and the strong coupling constant, 
as we have already mentioned before. Hence they can play an important role
for fine-tuning as it is discussed, for instance, in \cite{hep-ph/9303291}.   
Following the discussions there we weight the individual fine-tuning in this quantities
by their experimental uncertainty. It turns out that the fine-tuning in the strong coupling
constant is then very small and only the fine-tuning in $y_t$ plays a role. Due to the
rather complicated dependence of the coefficients on these two parameters we will
only include their effects in the numerical results. 
For the semianalytical results we have fixed $y_t$ and $\alpha_s$ to their best-fit values,
and varying them within their experimentally allowed ranges 
does not change the results qualitatively.

In the light of the recent Higgs results the CMSSM unavoidably has a certain amount 
of fine-tuning either from quite heavy squarks or from a large $A$-term (in maximal mixing scenarios).   
In this paper, we therefore consider the possibility that the underlying theory predicts different, non-universal 
(but fixed) boundary conditions at the GUT scale, namely non-universal scalar masses (NUSM) and 
non-universal gaugino masses (NUGM). 
Although such relations may not hold exactly 
in realistic models, they may guide towards more natural SUSY scenarios.

\subsection{Fine-Tuning from the Scalar Sector}

We now discuss possible NUSM scenarios.
To start with, let us note that the situation of universal sfermion masses in the CMSSM features a significant reduction of fine-tuning, since an almost-complete cancellation occurs automatically for the largest contributions in Eq.\ (\ref{eq:mHuSq}), i.e.\ between the term with $m_{h_u}^2$ and the terms with $m_{\tilde Q_3}^2$ and $m_{\tilde U_3}^2$. From the point of view of naturalness, such a fixed relation between $m_{h_u}^2$ and $m_{\tilde Q_3}^2, m_{\tilde U_3}^2$ is desirable.

Let us now consider non-universal fixed ratios for
the soft scalar masses motivated by unified theories:
if we assume that at the GUT
scale the gauge interactions unify to one single
interaction then the soft masses for
the various fields are not independent anymore
because they are (partially) unified in common
representations. Two very prominent paths to
unification are $SU(5)$ \cite{Georgi:1974sy} and
Pati--Salam \cite{Pati:1974yy}, which can both be
embedded in $SO(10)$ \cite{Georgi:1974my}. For simplicity we assume that
possible higher order GUT symmetry breaking corrections,
which might induce splittings within one representation
are negligibly small and we remind that we have set the first two
generations to have the same soft SUSY breaking masses.

\begin{itemize}
\item We begin with the case of $SU(5)$, where we find for $m^2_{H_u}$
\begin{equation}
\begin{split}
 \label{eq:mHuSqScalarsSU5}
 m^2_{H_u} (M_{\text{SUSY}}) &=
- 0.0012 m_{\tilde{T}_1}^2 + 0.0019 m_{\tilde{F}_1}^2
- 0.6083 m_{\tilde{T}_3}^2 + 0.0091 m_{\tilde{F}_3}^2 \\
&\quad + 0.6481 m_{h_u}^2 + 0.0273 m_{h_d}^2 \\
&\quad + \text{gaugino masses and trilinear terms} \;,
\end{split}
\end{equation}
where $m_{\tilde{T}_i}^2$ stands for the masses of the tenplet of $SU(5)$,
which contains the coloured doublet, the up-type coloured singlet and charged
leptonic singlet ($m_{\tilde{Q}_i}^2=m_{\tilde{U}_i}^2=m_{\tilde{E}_i}^2=m_{\tilde{T}_i}^2$), and $m_{\tilde{F}_i}^2$ for the fiveplet of $SU(5)$, which
contains the down-type coloured singlet and the leptonic doublet ($m_{\tilde{D}_i}^2=m_{\tilde{L}_i}^2=m_{\tilde{F}_i}^2$). We allow an
arbitrary splitting from the Higgs fields from the other scalar mass parameters
and also among each other. One can clearly see that
the simplest option
in order not to strongly increase the fine-tuning
is to impose 
a fixed relation where $m_{\tilde{T}_3}^2 \approx m_{h_u}^2$ holds, otherwise one has to consider the fine-tuning from each parameter separately, cf.\ Eq.\
\eqref{eq:wholetuning}. 

\item We come now to the Pati--Salam case for which we find
\begin{equation}
\begin{split}
 m^2_{H_u} (M_{\text{SUSY}}) &=
0.0029 m_{\tilde{l}_1}^2 - 0.0022 m_{\tilde{r}_1}^2 - 0.3102 m_{\tilde{l}_3}^2 - 0.2890 m_{\tilde{r}_3}^2 + 0.6754 m_{h}^2  \\
&\quad + \text{gaugino masses and trilinear terms} \;,
\end{split}
\end{equation}
where in this case $\tilde{l}$ denotes the left-handed doublets ($m_{\tilde{Q}_1}^2=m_{\tilde{L}_i}^2=m_{\tilde{l}_i}^2$), $\tilde{r}$
denotes the right-handed doublets ($m_{\tilde{U}_1}^2=m_{\tilde{D}_i}^2=m_{\tilde{E}_1}^2=m_{\tilde{r}_i}^2$) and $h$ denotes the Higgs bi-doublet ($m_{h_u}^2=m_{h_d}^2=m_{h}^2$). The
conclusions are similar to the $SU(5)$ case. The simplest fixed relation one
can
impose is 
$m_{\tilde{l}_3}^2 \approx m_{\tilde{r}_3}^2 \approx m_{h}^2$.

\item Finally, let us discuss the situation in $SO(10)$ GUTs.
Here one considers the soft mass terms for the matter fields
$m_{16_1}^2$ and $m_{16_3}^2$ and for the Higgs fields
$m_{h_u}^2=m_{h_d}^2=m_{10}^2$
(with possible $m_{h_u}^2 \not= m_{h_d}^2$ from D-term splitting),
obtaining
\begin{equation}
\begin{split}
 m^2_{H_u} (M_{\text{SUSY}}) &=
- 0.0031 m_{16_1}^2 - 0.5992 m_{16_3}^2 + 0.6754 m_{10}^2 \\
&\quad + \text{gaugino masses, trilinear terms and D-term splitting terms} \;,
\end{split}
\end{equation}
which implies that again, a fixed relation beyond the ones from the GUT itself would have to be imposed (e.g.\ between $m^2_{10}$ and $m^2_{16_3}$) in order to strongly decrease fine-tuning.

\end{itemize}

A few additional comments are in order.
To start with, one can see that the soft terms for the first two families enter with small coefficients into $m^2_{H_u}$. This illustrates the well-known fact that these soft masses can be significantly larger 
than the third family ones, without paying a large fine-tuning price. Furthermore, in the above discussion we have focused on fixed relations from GUT structures. As we have already noted, with these fixed relations alone one would increase fine-tuning compared to the CMSSM case. From the above equations one can imagine additional fixed relations on top of these structures which could in principle reduce fine-tuning by effectively leading to a cancellation in the contributions to $m^2_{H_u}$. However, we will not investigate such cases in more detail here. 

Another relevant remark is that we have only discussed here ``direct'' effects on the fine-tuning, i.e.\ the effects on the fine-tuning measure for a fixed SUSY parameter point. We note that there are also ``indirect'' effects expected, i.e.\ when introducing a non-universality allows to avoid certain constraints on the SUSY parameter space and makes regions with lower fine-tuning accessible (e.g.~a ``compressed'' 
spectrum helps to evade LHC constraints \cite{Arbey:2011un}). 

Finally, let us notice that small splittings in models where the soft parameters are universal (CMSSM-like) at the leading order and get only corrected by small additional contributions (e.g.\ by higher dimensional operators in flavour models, see, for instance, \cite{Antusch:2011sq}) can have a certain degree of non-universality without increasing the fine-tuning too much. However, large non-universalities ($\mathcal{O}(1)$) again lead back to the same situation as in the general case \'{a} la pMSSM.

From the above equations one can imagine 
additional fixed relations but the simplest possibility
is to have the scalar mass parameters -- at least the
high energy stop and Higgs soft masses -- almost
degenerate to keep the fine-tuning small, like for
instance in the CMSSM. Non-universal scalar
masses tend to make the fine-tuning worse rather
than improving the situation.
We will therefore not study this case numerically in more detail.

\subsection{Fine-Tuning from Gaugino Masses and Trilinear Couplings}

Now we turn to the gaugino masses and the trilinear SUSY
breaking couplings.
They appear together in the $m^2_{H_u}$ polynomial
and hence should not be discussed independently.
For the sake of simplicity we assume here that the trilinear
couplings are universal
$\eta_t = \eta_b = \eta_\tau = \eta_A$.\footnote{
In principle, one can reduce the fine-tuning as well by
assuming definite relations between $\eta_t$, $\eta_b$, and $\eta_\tau$
just as we will demonstrate here for the gaugino mass parameters.
However, the coefficients involving $\eta_b$ and $\eta_\tau$ in
Eq.\ \eqref{eq:mHuSq} are very small and hence the values of $\eta_b$ and $\eta_\tau$ should be very large.
}

In this case the $m^2_{H_u}$ polynomial reads
\begin{align}
  m^2_{H_u} (M_{\text{SUSY}}) &=
   - M_3^2 (1.2865 - 0.0216 \eta_1 - 0.0242 \eta_1^2 + 0.0230 \eta_2 - 0.2177 \eta_2^2 + 0.0813 \eta_1 \eta_2 ) \nonumber\\
 & \quad + M_3^2 \eta_A (0.2904 + 0.0112 \eta_1 + 0.0652 \eta_2) - 0.1131 M_3^2 \eta_A^2 \nonumber\\
 & \quad + \text{scalar masses}  \nonumber\\
 & \equiv (f_1 (\eta_1, \eta_2) + f_2(\eta_1,\eta_2) \eta_A + f_3 \eta_A^2 ) M_3^2 + \ldots \;, \label{eq:FTGauginos}
\end{align}
where we have again set $\eta_3 = 1$.
Now the question arises for which values of the parameters 
the fine-tuning, as defined in Eqs.~\eqref{eq:FT_mi_pMSSM}
and \eqref{eq:FT_mu_pMSSM}, is minimised.

\begin{figure}
\centering
\includegraphics[scale=0.5]{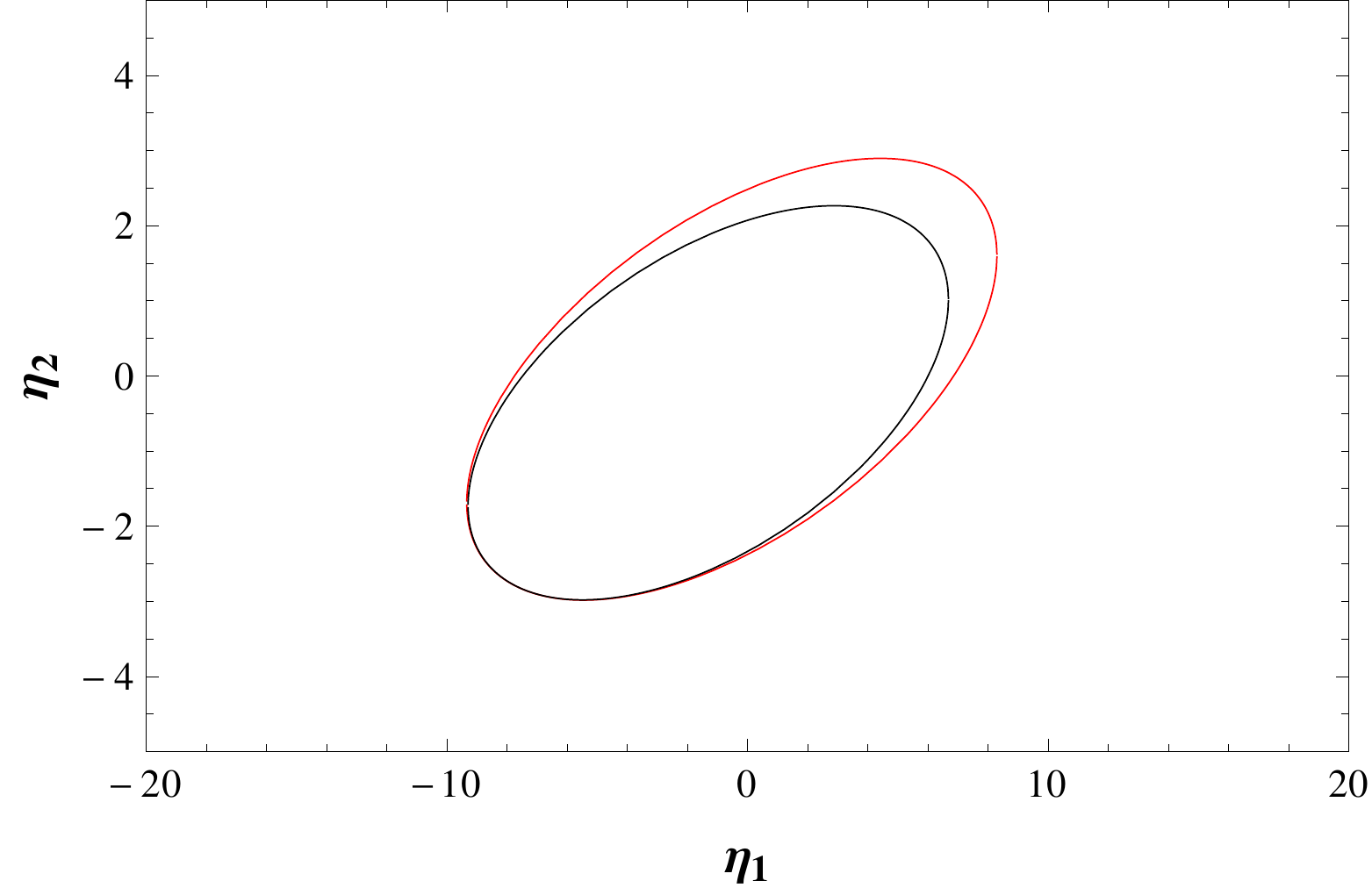}
\caption{ \label{fig:ellipses}
The two ellipses in the $\eta_1$-$\eta_2$
plane which minimise the fine-tuning. The red
ellipse corresponds to Eq.\ \eqref{eq:Sol1}
and the black one to Eq.\ \eqref{eq:Sol2}.
}
\end{figure}

The overall fine-tuning, considering only this sector is
\begin{equation}
 \Delta \approx \left| \frac{M_3^2}{M_Z^2} \right| \max \begin{Bmatrix}
 |2 f_3 \eta_A^2 + f_2(\eta_1, \eta_2) \eta_A|, \\
 |2 \, f_1(\eta_1, \eta_2) + f_2(\eta_1, \eta_2) \eta_A|
 \end{Bmatrix} \;.
\end{equation}
Setting these two equations to zero we find two possible solutions:
\begin{align}
    \text{\bf Solution 1: }& f_1 (\eta_1, \eta_2) = 0, & \eta_A &= 0 \;, \label{eq:Sol1}\\
    \text{\bf Solution 2: }& f_1 (\eta_1, \eta_2) = \frac{f_2 (\eta_1, \eta_2)^2}{4 f_3}, & \eta_A &= -\frac{f_2 (\eta_1, \eta_2)}{2 f_3} \;. \label{eq:Sol2}
\end{align}
With the parameters in Eq.\ \eqref{eq:FTGauginos}, the two solutions define two different ellipses in the $\eta_1$-$\eta_2$ plane,
see Fig.\ \ref{fig:ellipses}. As we will see in our
numerical analysis later on this is a generic feature
of the parameter space.

As the A-terms are crucial for the prediction of the physical Higgs mass, 
it is interesting to discuss
the dependence on $\eta_A$ in more detail. For the first solution
this is trivial, while the second solution can be written
in terms of $\eta_1$ and $\eta_2$ depending on $\eta_A$
so that we find 
\begin{equation}
\begin{split} \label{eq:ellipse}
 \eta_1 &= -8.0849 + 6.0757 \eta_A \mp 4.9510 \sqrt{(2.1354 - \eta_A)(\eta_A - 0.0962)} \;, \\
 \eta_2 &= -3.0681 + 2.4288 \eta_A \pm 0.8483 \sqrt{(2.1354 - \eta_A)(\eta_A - 0.0962)} \;,
\end{split}
\end{equation}
which gives real ratios only for $0.0962 < \eta_A < 2.1354$.
Interestingly, this implies for example that in our benchmark
point with $\eta_A = -5$ the fine-tuning cannot be made arbitrarily
small by choosing appropriate $\eta_1$ and $\eta_2$.
Still the fine-tuning can be significantly reduced:
for this benchmark point we find a minimal fine-tuning
in the gaugino sector of
\begin{equation}
 \Delta = 4.95 \left| \frac{M_3^2}{M_Z^2} \right| \;,
\end{equation}
for $\eta_1 = -11.7$ and $\eta_2 = -4.6$ compared
to $\Delta = 7.49 |M_3^2/M_Z^2|$ for
$\eta_1 = \eta_2 =1$.

Note that we have discussed here the fine-tuning
in $M_3$ and $\eta_A$. As we mentioned above, there are regions of the
parameter space where the fine-tuning in $\mu$ is dominant. 
Nevertheless, $\Delta_\mu$
is also reduced by the choices of $\eta_1$, $\eta_2$
and $\eta_A$ we have just discussed.
From Eq.\ \eqref{eq:FTmu} we see that
making $|m_{h_u}^2|$ small reduces the fine-tuning
in $\mu$.
For solution 1 this is obvious. In the case of solution 2,
it is easy to check that the contribution to $|m_{h_u}^2|$
from the gaugino masses and trilinears is set to
zero, too.

The first solution for the case $\eta_A = 0$ was already found in \cite{Horton:2009ed} - albeit with a slightly different notation setting $\eta_2 = 1$ instead of $\eta_3 = 1$ and thus giving hyperbolas instead of ellipses. The second solution, however, was not discussed therein. Instead, in \cite{Abe:2007kf} a related case was studied where also a fixed relation between the gaugino and the trilinear mass scales was assumed, while in our analysis $\eta_A$ is a free parameter. We will discuss which GUT scale gaugino mass ratios are favoured, comparing numerical with semi-analytical results, in section~\ref{sec:num}.

\section{GUT Scale Yukawa Ratios and SUSY Threshold Corrections}

In GUTs quarks and leptons are unified in representations of the GUT symmetry group, which implies that the elements of the Yukawa matrices are generally related. As a consequence, fixed ratios of quark and lepton masses are typically predicted by GUT models. These ratios are given by group theoretical Clebsch--Gordan factors and hold at the GUT scale. For the third family, phenomenologically interesting possibilities are, e.g., the standard bottom-tau Yukawa unification $y_\tau / y_b = 1$ or the recently proposed ratio $y_\tau / y_b = 3/2$ \cite{arXiv:0902.4644}. The GUT scale Yukawa ratios are thus interesting discriminators between GUT models. We are therefore also interested in which GUT-scale Yukawa coupling relations might be preferred due to lower fine-tuning.

The phenomenological viability of GUT scale Yukawa ratios depends on SUSY
threshold corrections \cite{SUSYthresholds}.
Here we give some short comments on this important class of one-loop corrections
to the Yukawa couplings of the down-type quarks and the charged
leptons in the MSSM, which are sizeable if $\tan \beta$ is large. 
These corrections depend on the SUSY breaking parameters
and hence the GUT scale Yukawa coupling ratios depend on the
SUSY spectrum.

For example for the bottom quark mass we can write
\begin{equation}
m_b = y_b v_d (1 + \epsilon_b \tan \beta) \;,
\end{equation}
the correction $\epsilon_b$ being given by \cite{Freitas:2007dp, Antusch:2008tf,Spinrath:2010dh}
\begin{equation} \label{eq:epsilonb}
\epsilon_b \approx \epsilon^G + \epsilon^B + \epsilon^W + \epsilon^y \;,
\end{equation}
where
\begin{align} \label{eq:expq}
 \epsilon^G & =  -\frac{2 \alpha_S}{3 \pi} \frac{\mu}{M_3} H_2(u_{\tilde{Q}_3},u_{\tilde{d}_3}) \; , \\
\label{eq:expq_bino}
 \epsilon^B & =  \frac{1}{16 \pi^2} \left[ \frac{{g'}^2}{6} \frac{\eta_1 M_3}{\mu} \left( H_2(v_{\tilde{Q}_3}, x_1) + 2 H_2(v_{\tilde{d}_i}, x_1) \right) + \frac{{g'}^2}{9} \frac{\mu}{\eta_1 M_3} H_2(w_{\tilde{Q}_3},w_{\tilde{d}_3}) \right], \\
 \epsilon^W & =  \frac{1}{16 \pi^2} \frac{3 g^2}{2} \frac{\eta_2 M_3}{\mu} H_2(v_{\tilde{Q}_3}, x_2) \; , \\
 \epsilon^y & = - \frac{y_t^2}{16 \pi^2} \frac{\eta_t M_3}{\mu} H_2(v_{\tilde{Q}_3}, v_{\tilde{u}_3}) \; .
\label{eq:expq_end}
\end{align}
Here $u_{\tilde{f}} = m_{\tilde{f}}^2/M_3^2$, $v_{\tilde{f}} = m_{\tilde{f}}^2/\mu^2$, $w_{\tilde{f}} = m_{\tilde{f}}^2/(\eta_1^2 M_3^2)$, $x_i = (\eta_i^2 M_3^2)/\mu^2$ for $i=1,2$ and the loop function $H_2$ reads
\begin{equation}
    H_2(x, y) = \frac{x \ln x}{(1-x)(x-y)} + \frac{y \ln y}{(1-y)(y-x)} \;.
\end{equation}
In the CMSSM one can neglect in a first approximation $\epsilon^B$
and $\epsilon^W$ because they are suppressed by the small gauge couplings.
In NUGM scenarios this is in general not true anymore
because the suppression might be compensated by an enhancement of the
bino or wino mass parameter compared to the gluino one. In our
example point we have seen that, for example, $\eta_1 = -11.7$ and
$\eta_2 = -4.6$ is preferred to have low fine-tuning.

Without some knowledge of the parameter space or simplifying assumptions a
quantitative statement is hence quite difficult. But in general one can
expect significant corrections up to 50~\% for the GUT scale Yukawa
coupling ratios (see, e.g.~\cite{Antusch:2008tf,Spinrath:2010dh}).

\section{Numerical Analysis}
\label{sec:num}

To improve our understanding based on the semi-analytical treatment 
above we now turn to our numerical analysis. To this end we employed 
a modified version of {\tt softSUSY} \cite{Allanach:2001kg} to 
calculate the SUSY spectra, GUT scale Yukawa couplings and the fine-tuning. 
The modifications take into account the implicit $M_Z$ dependence 
on the Higgs vev \cite{Cassel:2009ps} and the SUSY threshold
corrections for the Yukawa couplings of all three fermion generations
\cite{SUSYthresholds, Freitas:2007dp, Antusch:2008tf,Spinrath:2010dh}.

In order to compare with experimental constraints the observables BR$(b \to s \gamma)$,
BR$(B_s \to \mu^+ \mu^-)$ and BR$(B_u \to \tau \nu_\tau)$ were calculated using {\tt SuperIso} \cite{SuperIso}.
The experimental ranges we considered for such observables are: 
BR$(b \to s \gamma) = (355 \pm 24 \pm 9)10^{-6}$ \cite{Asner:2010qj},
BR$(B_s \to \mu^+ \mu^-) < 4.5 \cdot 10^{-9}$ at $95\%$ CL \cite{Aaij:2012ac} and BR$(B_u \to \tau \nu_\tau) = 1.41 \pm 0.43$ \cite{Barberio:2008fa}. 
Additionally, LEP bounds \cite{PDG} on sparticle masses were applied
and we discard points corresponding to a $\tilde{\tau}$ LSP and color
and charge breaking (CCB) vacua.
We present results with and without the Higgs boson mass constraint
included to study its impact.
Applying constraints from direct LHC SUSY searches is not
straightforward and we will only show a comparison to simplified
models for illustration. 

In our opinion the $3.2 \sigma$ level $(g-2)_\mu$ discrepancy from the Standard Model expectation \cite{g-2} is an experimental evidence that, while certainly interesting, still requires a full experimental confirmation from next generation experiments, such as those proposed at J-PARC \cite{g-2exp_jparc} and at Fermilab \cite{g-2exp_fermilab}. For this reason we have not imposed constraints from the anomalous magnetic moment of the muon. In fact in this scenario we can only find very few and isolated points which would satisfy $(g-2)_\mu$ at 2$\sigma$ and predict a Higgs boson mass in agreement with the recent results, all of them corresponding to a fine-tuning larger than ${\cal O} (100)$.

Moreover, we have not imposed constraints from neutralino relic density nor from direct and indirect dark matter searches. Such constraints can be evaded assuming a non-standard cosmological history or a different
LSP, like the gravitino or the axino. 
Nevertheless, it is interesting to note that for $\eta_2 \lesssim 0.5\, \eta_1$, the lightest neutralino is dominated by its Wino (or Higgsino) component, which implies that the relic density is strongly suppressed. In these regions, on the one hand, one cannot explain dark matter by thermal relic neutralinos, but, on the other hand, there is no overproduction of neutralino dark matter.

\subsection{Before the latest LHC Higgs and SUSY Results}
\label{sec:num_before}

\begin{figure}
\centering
\includegraphics[width=0.6\textwidth]{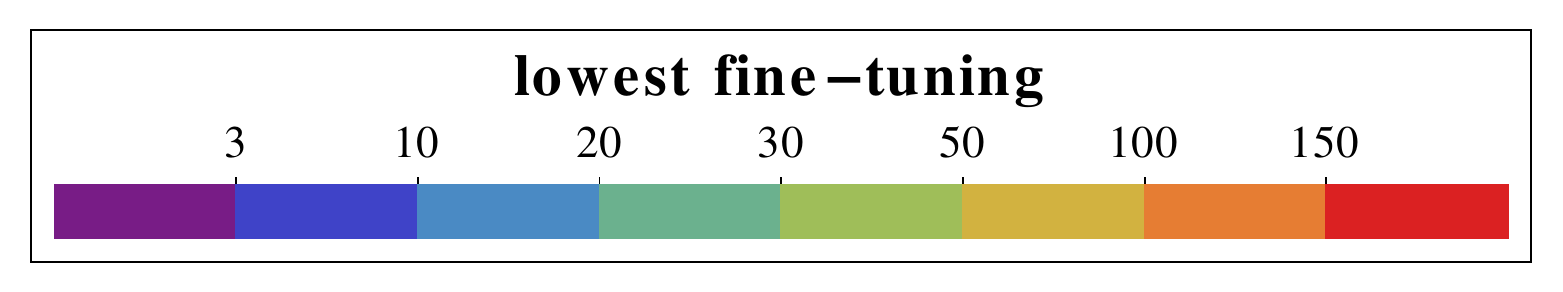}

\subfigure[]{\includegraphics[width=0.47\textwidth]{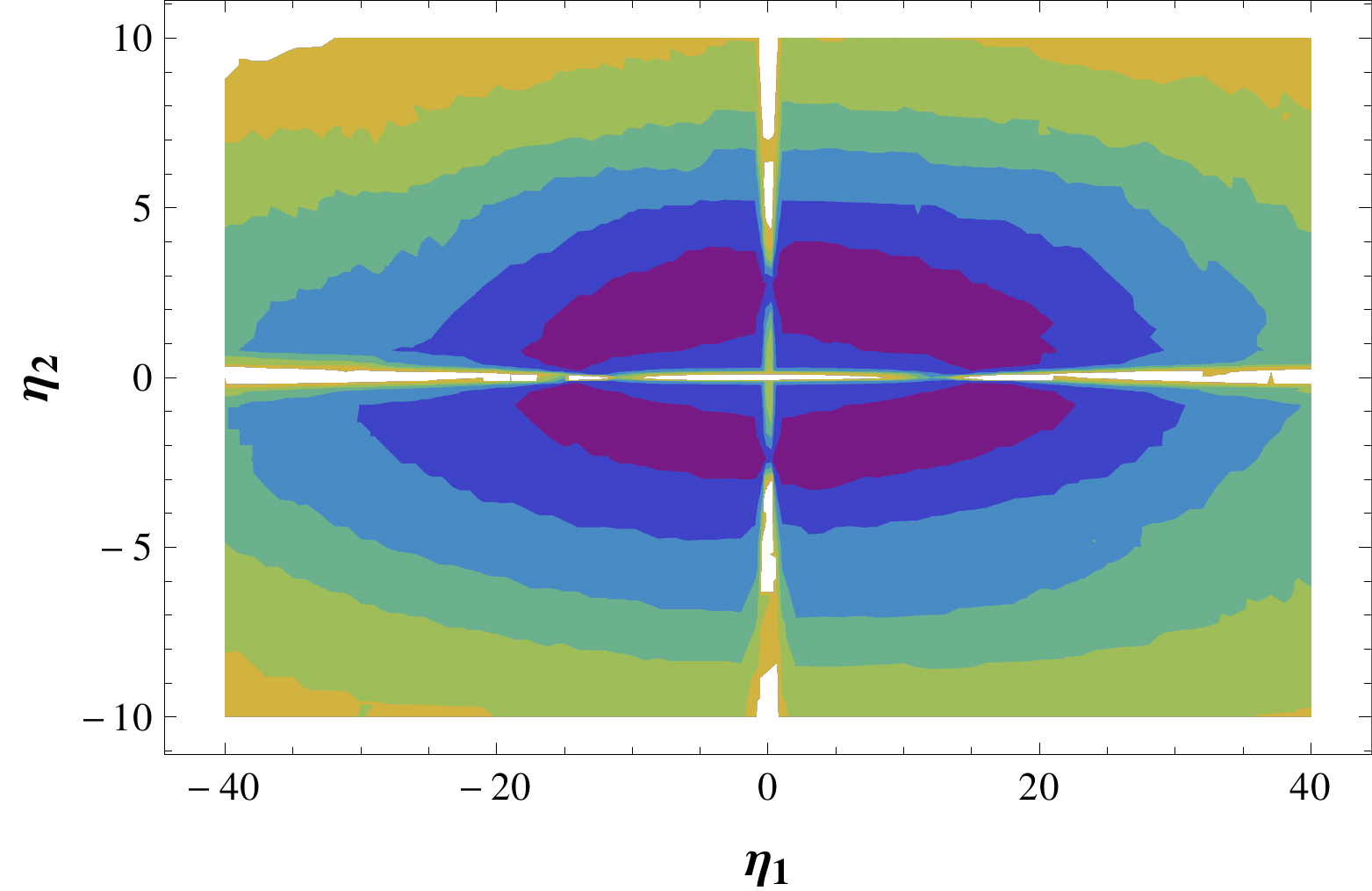}}
\subfigure[]{\includegraphics[width=0.47\textwidth]{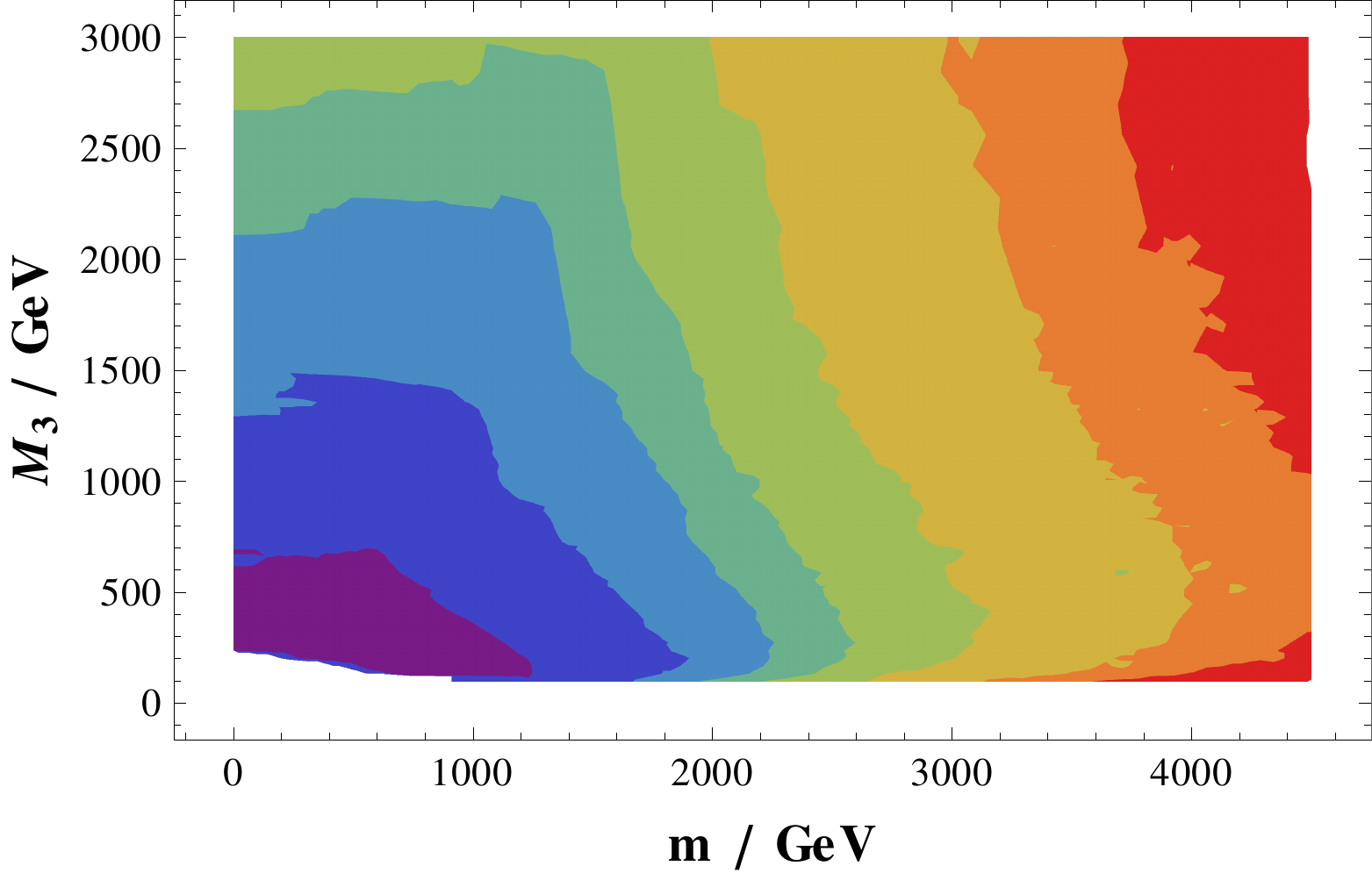}}

\subfigure[]{\includegraphics[width=0.47\textwidth]{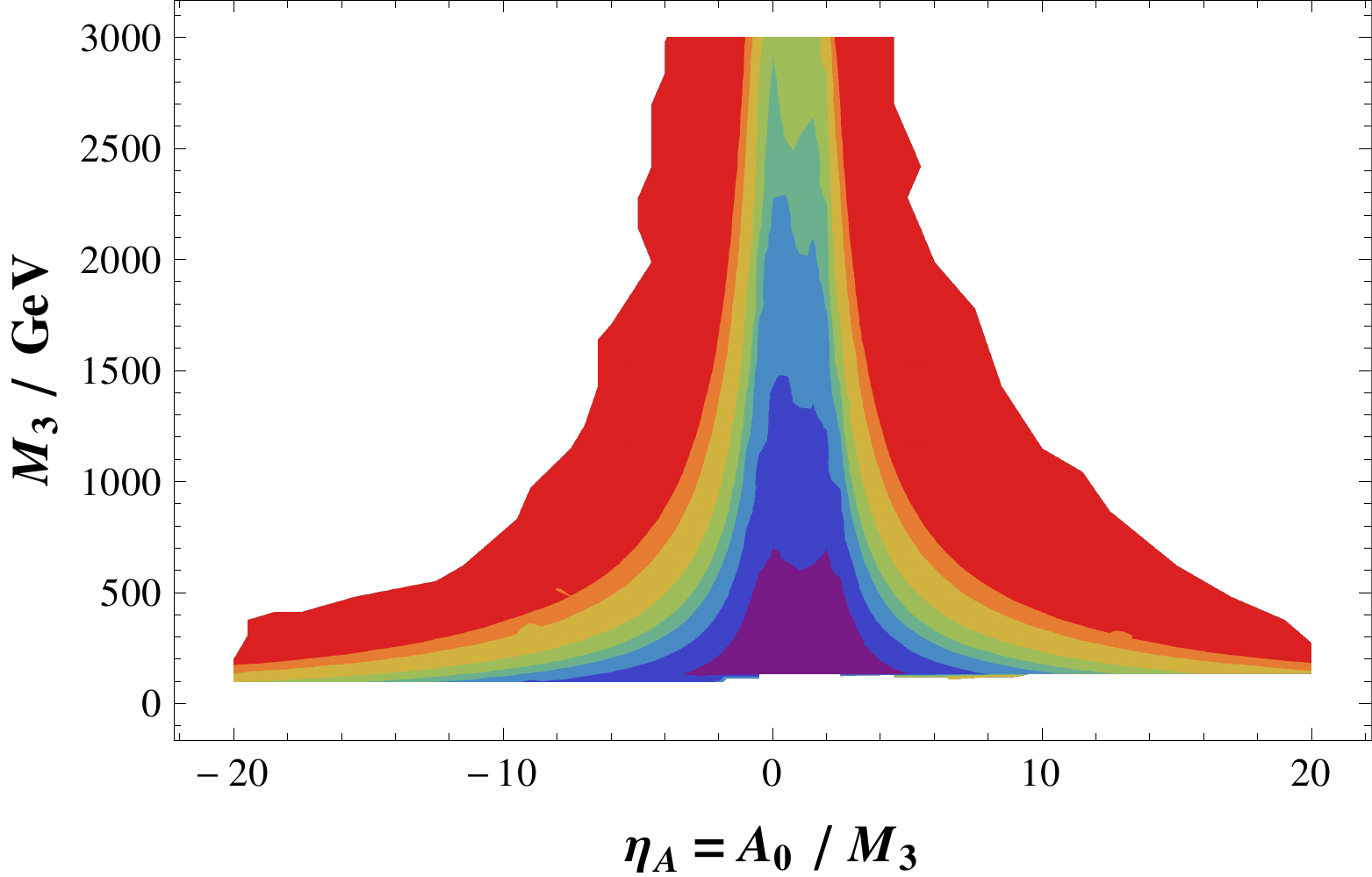}}
\subfigure[]{\includegraphics[width=0.47\textwidth]{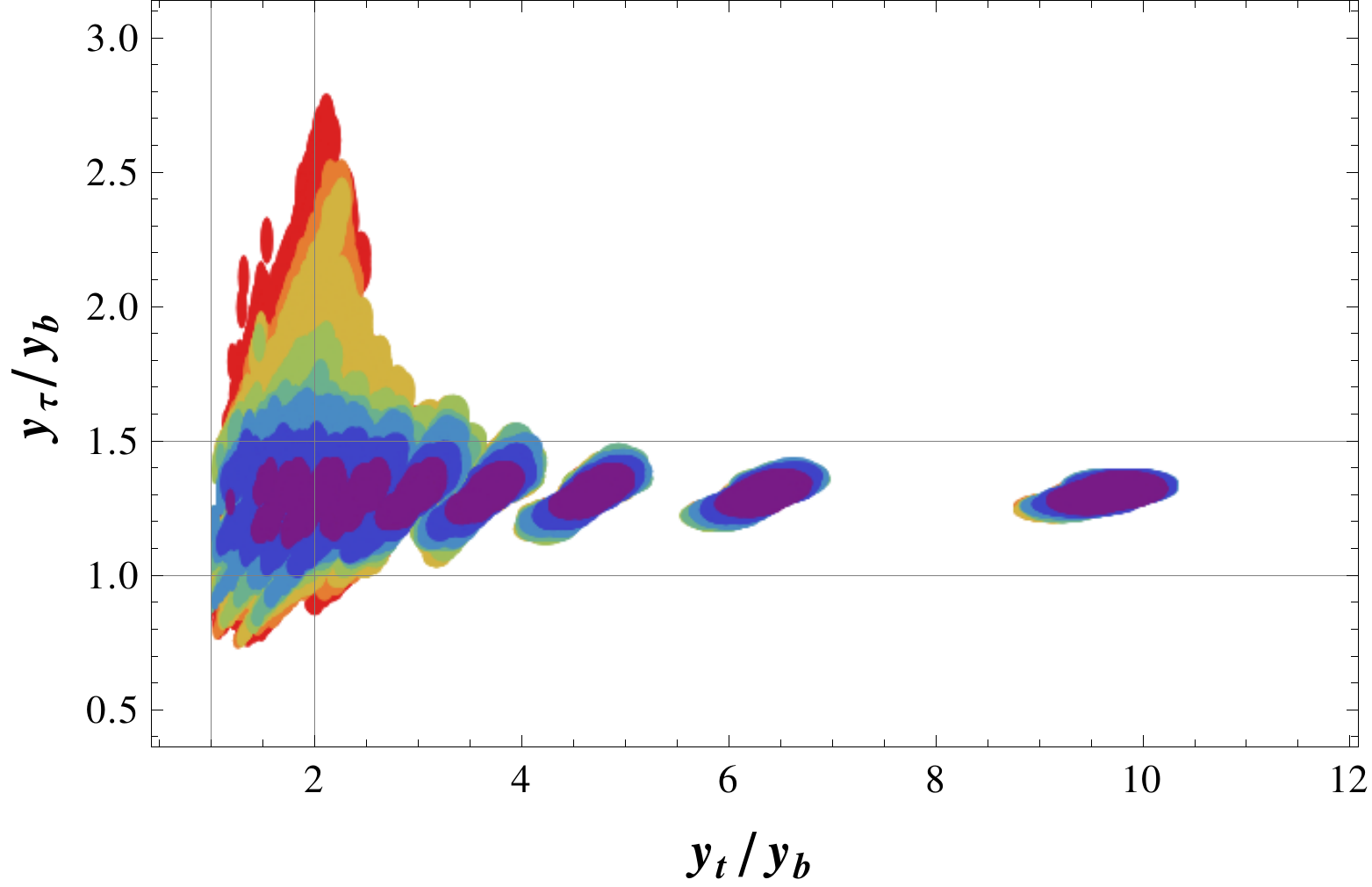}}
\caption{
Lowest fine-tuning in the (a) $\eta_1$-$\eta_2$, (b) $m$-$M_3$, (c) $\eta_A$-$M_3$ and (d) $y_t/y_b$-$y_\tau/y_b$  planes consistent with the experimental bounds described in the text. For explanations of the parameters see main text. The Higgs mass constraint is not included. The 1$\sigma$ errors on the quark masses \cite{PDG} are taken into account by scaling the data points in the last plot correspondingly.
}\label{fig:BeforeMh}
\end{figure}

As previously discussed, we expect to find an ellipse
shaped region in the $\eta_1$-$\eta_2$ plane where fine-tuning
is significantly lower than in other parts of the parameter space,
especially when compared to the CMSSM. To study this in detail, we
scanned the NUGM parameter space in the following region:
the soft scalar mass parameters were assumed to be universal
$m_i = m$ for every $i$ and varied from 0 to 4.5~TeV, the GUT
scale gluino mass $M_3$ from 0.15 to 2~TeV, and
$\eta_t=\eta_b=\eta_\tau= \eta_A$ from $-20$ to 20 for
$\tan \beta = 2$, 10, 15, 20, $\dots$, 60. 
For the ranges of the gaugino mass ratios we took
$-40  \leq \eta_1 \leq 45 $ and  $-10\leq \eta_2 \leq 10$.
Both signs for $\mu$ were allowed. In the scan we have dropped
all points with a fine-tuning larger than 200.

Fine-tuning was calculated for the parameters $m$, $\mu$, $A_0 = \eta_A M_3$ and $M_3$.
Nominally existing tuning in the ratios $\eta_1$, $\eta_2$ was neglected as we assume such relations to be rigid and given by some underlying model.
We note as a side remark that in all plots where contours are shown instead of only points, outliers that are very far away from their neighbours are still shown as isolated points instead of being incorporated into the contour. In addition, unless stated otherwise, all plots show values marginalised over parameters not shown.

As we can see from Fig.\ \ref{fig:BeforeMh}a, we indeed find an ellipse that allows for quite low fine-tuning and evades the experimental bounds.
We see that the semi-analytical results from Fig.\ \ref{fig:ellipses} do not very well reproduce the numerical results but still they give a good qualitative understanding.
Compared to the CMSSM with the same experimental bounds, we find a fine-tuning $\Delta$ lower than 3 instead of just below 10 \cite{Antusch:2011xz}.
In this section, we will focus on this region of low fine-tuning and only consider points with an $(\eta_1, \eta_2)$ combination lying on the ellipse defined by $\Delta_{\text{min}} < 10$.\footnote{Depending on the other parameters these points can have a larger fine-tuning but, interestingly, this does not significantly change the appearance of all affected plots, indicating that being on the ellipse is indeed a necessary condition for small fine-tuning.} 
An approximate analytical formula for the ellipse in Fig.\ \ref{fig:BeforeMh}a is given by
\begin{equation}\label{eq:rough_ellipse}
(\eta_1/15.0)^2 + (\eta_2/2.6)^2 = 1\:.
\end{equation}

Based on this, we take a closer look at the fine-tuning dependence on the other parameters. In Fig.\ \ref{fig:BeforeMh}b we can clearly see that introducing non-universal gaugino masses can significantly weaken the dependence of $\Delta_{\text{min}}$ on the gaugino mass scale compared to the CMSSM. While in the latter a fine-tuning of 20 is only possible for $M_3 < 400$ GeV~\cite{Antusch:2011xz}, now we can go up to $M_3 \sim 2.2$~TeV. On the other hand, the fine-tuning dependence on $m$ (resp.\ $m_0$ in the CMSSM) is very much the same as expected.

Furthermore from Fig.\ \ref{fig:BeforeMh}c we can roughly see the same behaviour for large $\eta_A$ as in the CMSSM~\cite{Antusch:2011xz}, while for small $\eta_A$ from -2 to 4 we find two peaks. They lie at $\eta_A = 0$ and $\eta_A \sim 2$, which are the values for $\eta_A$ required for solution 1 and the upper end of the range quoted for solution 2, cf.~Eqs.~(\ref{eq:Sol1}, \ref{eq:Sol2}). 

Finally, concerning the GUT scale Yukawa coupling ratios shown in Fig.~\ref{fig:BeforeMh}d we find no qualitative difference compared to the CMSSM, besides the rather obvious fact that the fine-tuning price of all ratios is greatly reduced. The comparison of the $b$-$\tau$ Yukawa coupling ratio of $3/2$ with the case of $b$-$\tau$ Yukawa unification, however, is slightly in favor of $3/2$ with a minimal fine-tuning of $\Delta \sim 5$ vs.\ $9$.

\subsection{Results including the latest LHC Higgs and SUSY Searches}

\begin{figure}
\centering
\includegraphics[width=0.6\textwidth]{plots/NUGM-legend.pdf}

\subfigure[]{\includegraphics[width=0.47\textwidth]{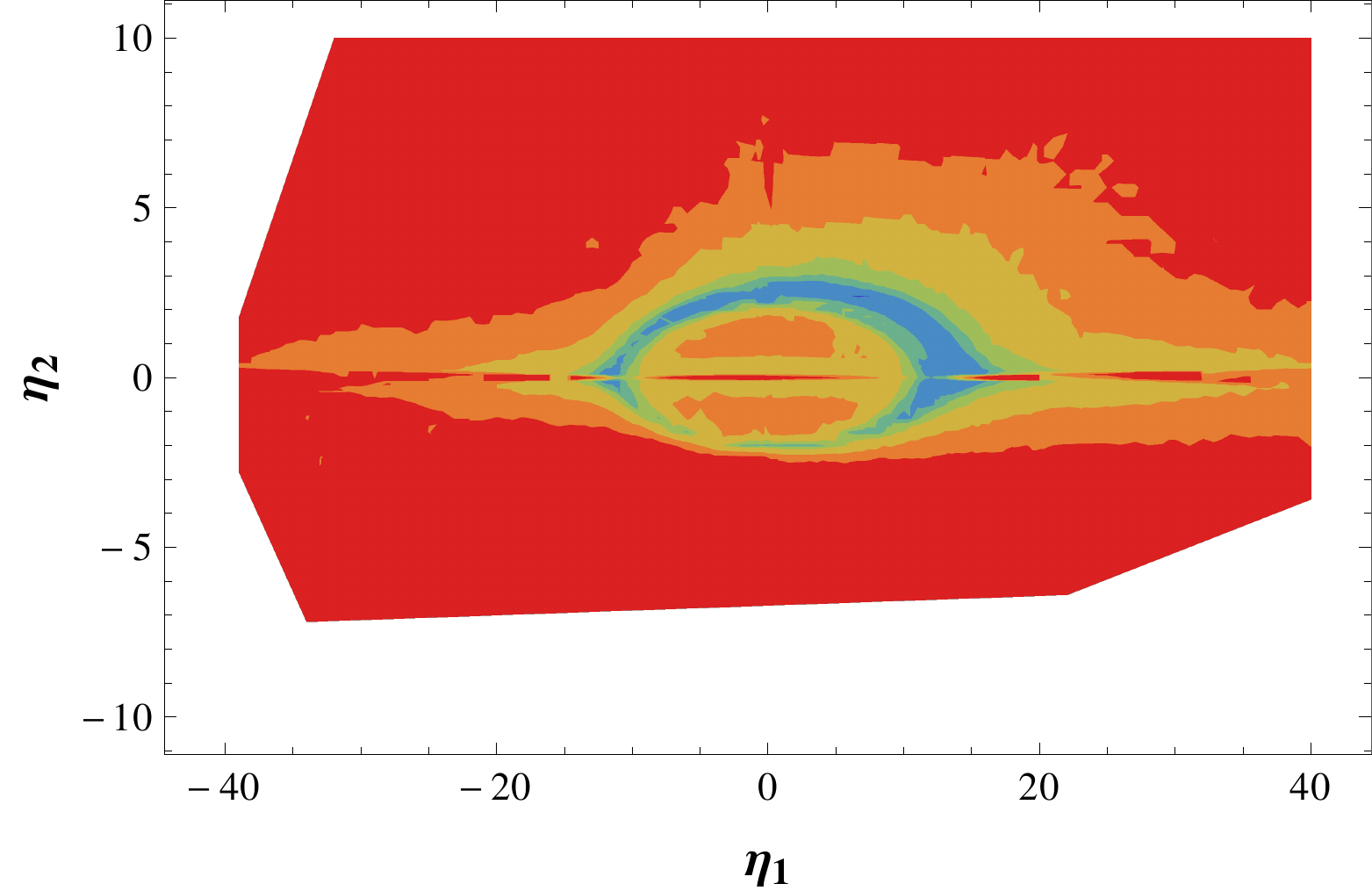}}
\subfigure[]{\includegraphics[width=0.47\textwidth]{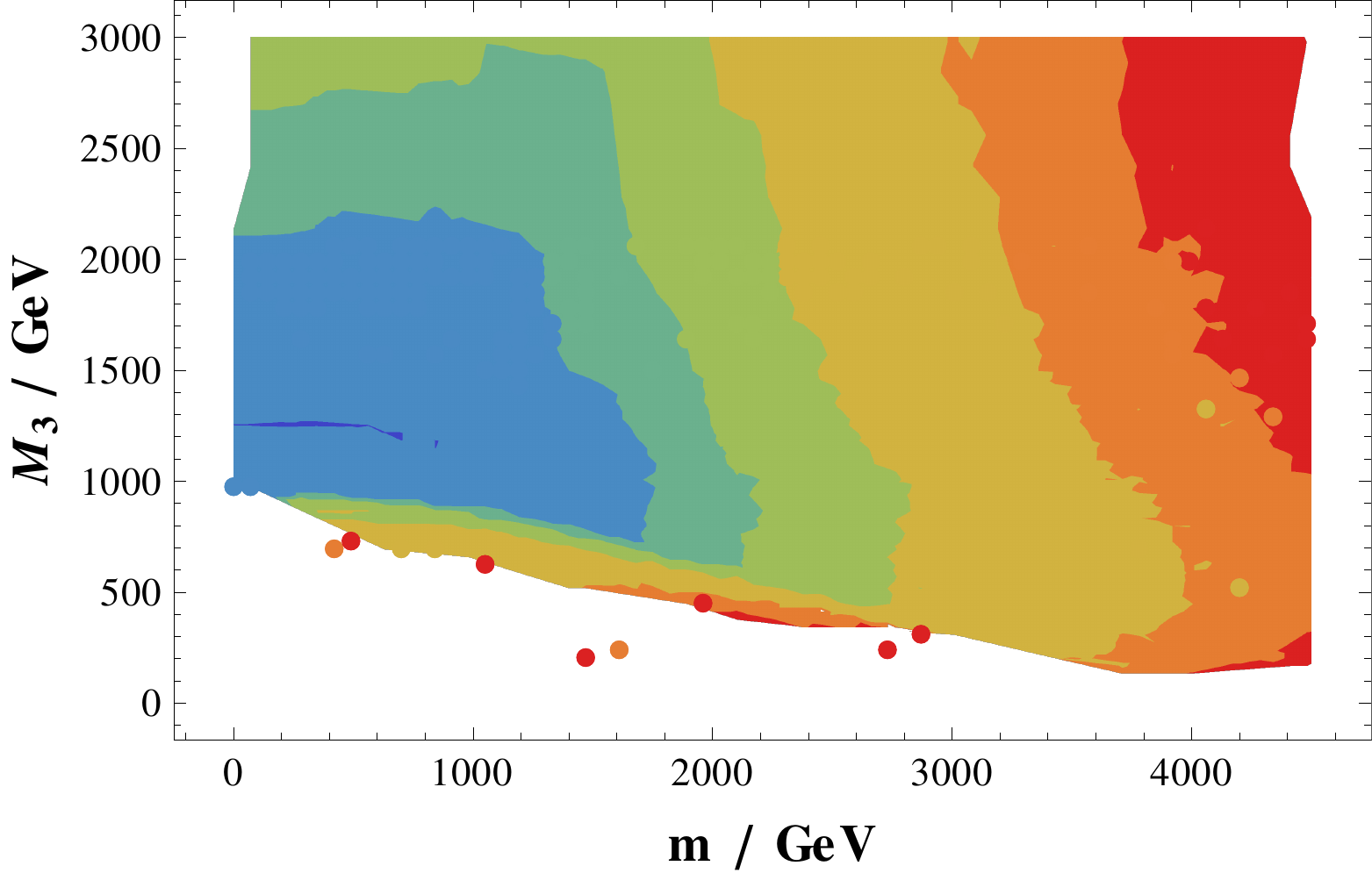}}

\subfigure[]{\includegraphics[width=0.47\textwidth]{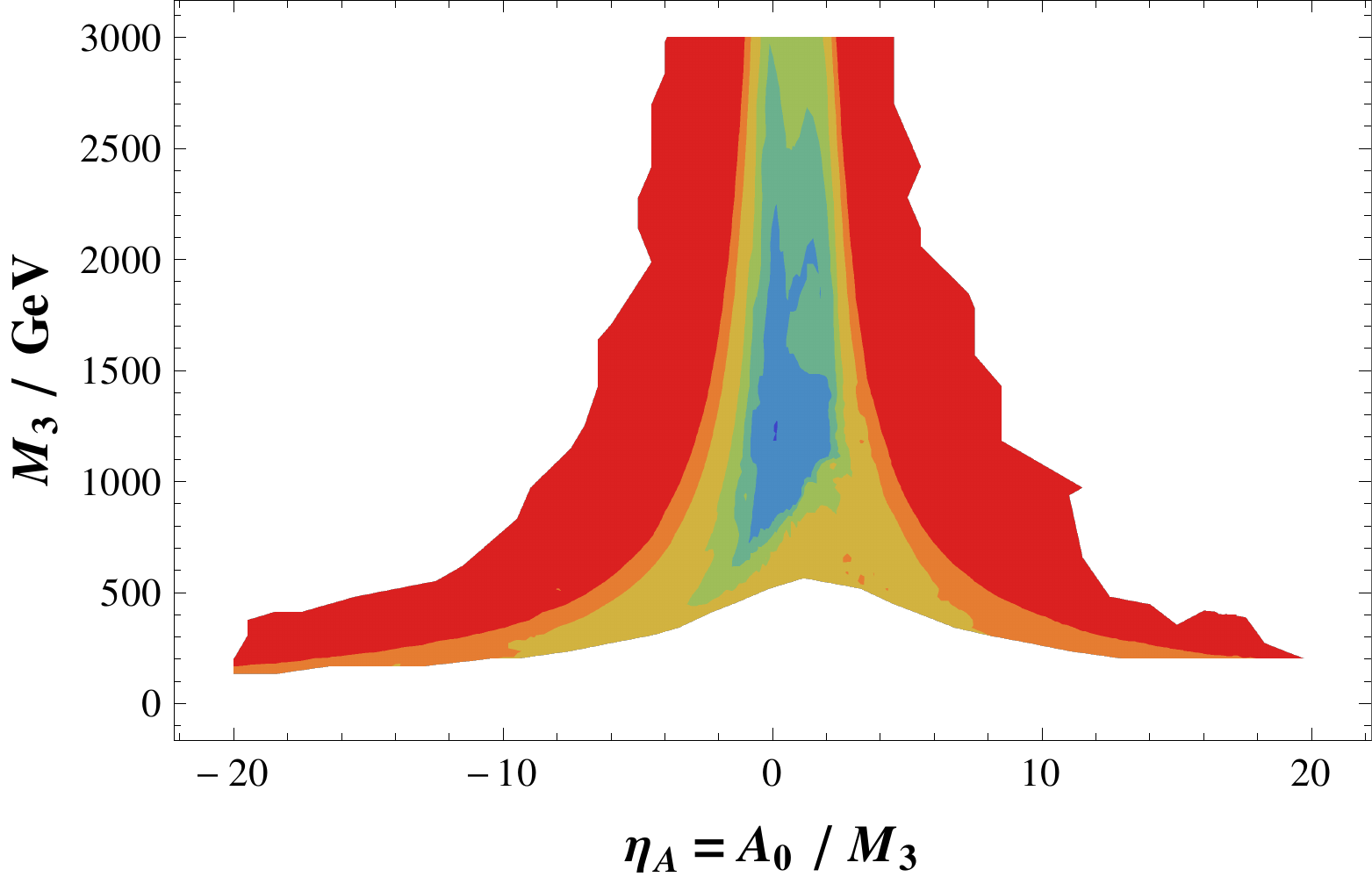}}
\subfigure[]{\includegraphics[width=0.47\textwidth]{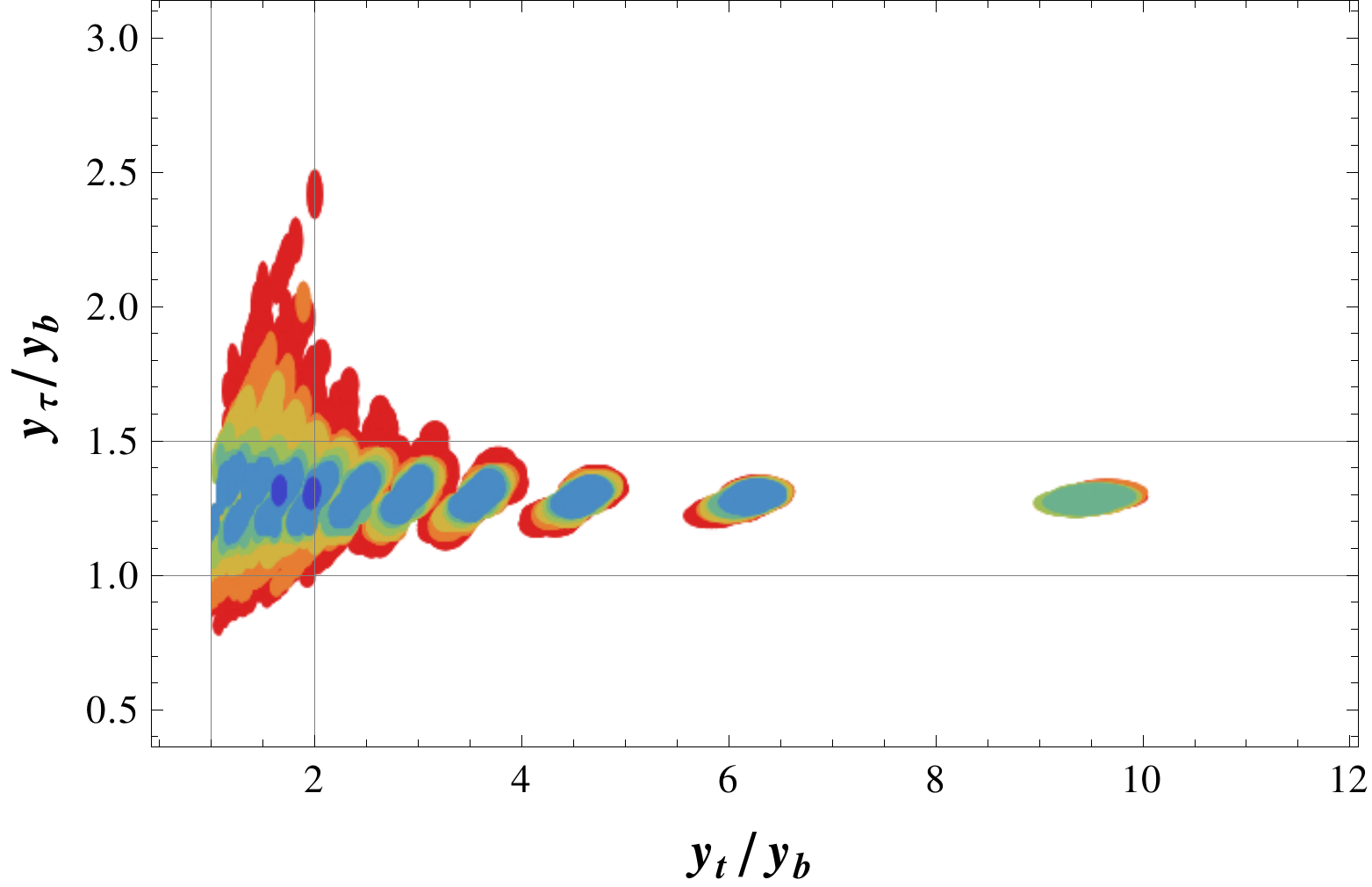}}
\caption{
Lowest fine-tuning in the (a) $\eta_1$-$\eta_2$, (b) $m$-$M_3$, (c) $\eta_A$-$M_3$ and (d) $y_t/y_b$-$y_\tau/y_b$  planes consistent with the experimental bounds described in the text. For explanations of the parameters see main text. In comparison to Fig.\ \ref{fig:BeforeMh} we have included the CMS experimental constraint $m_h = 125.3 \pm 0.6 $ GeV \cite{discovery} and a theoretical uncertainty of $\pm 3$ GeV \cite{Arbey:2012dq} for the Higgs mass calculation at each data point. The 1$\sigma$ errors on the quark masses \cite{PDG} are taken into account by scaling the data points in the last plot correspondingly.
}\label{fig:AfterMh}
\end{figure}

\begin{figure}
\centering
\includegraphics[width=0.47\textwidth]{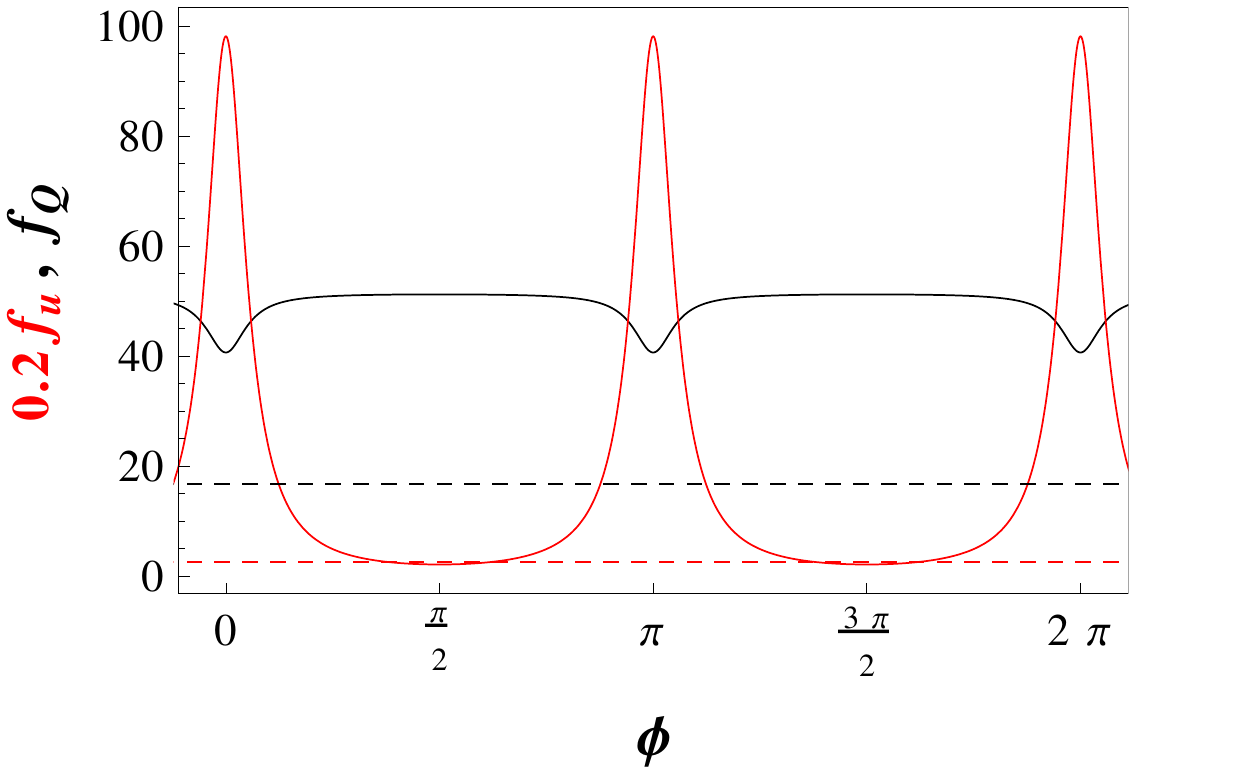}
\includegraphics[width=0.47\textwidth]{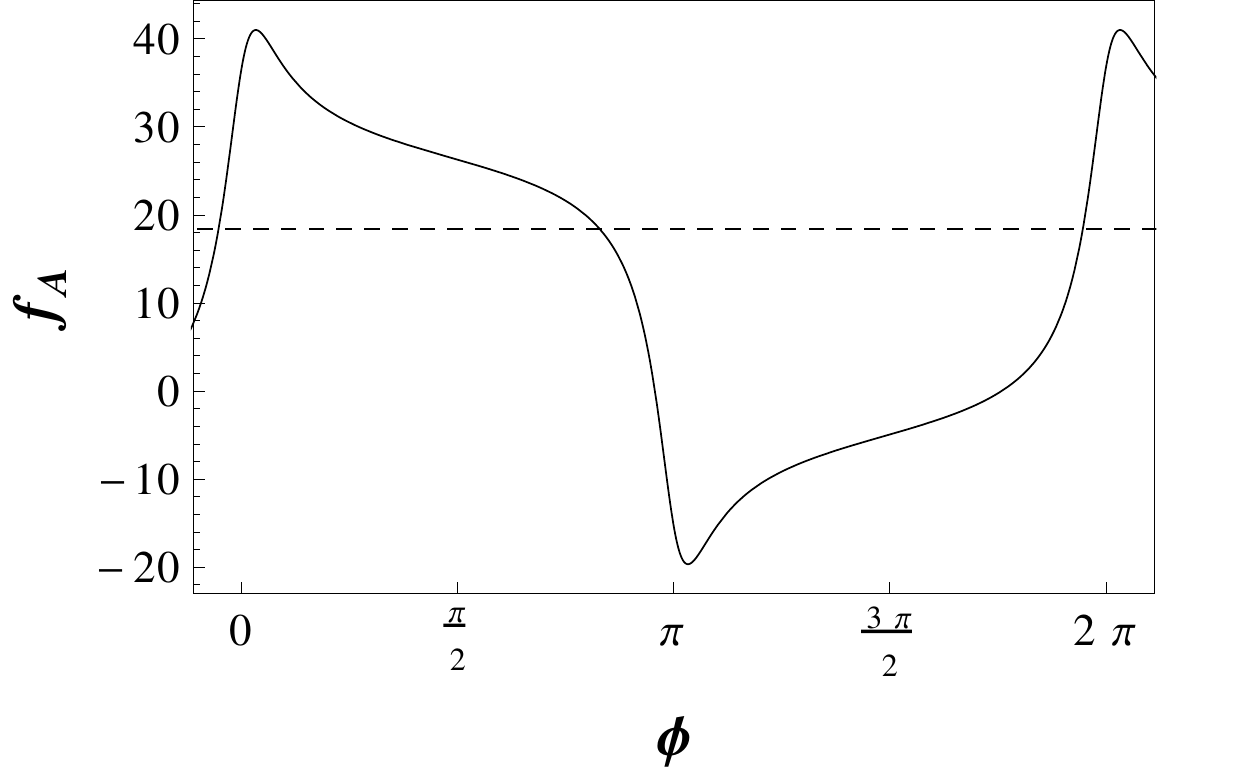}
\caption{
Dependence of gaugino mass contribution functions $f_Q$, $f_u$, $f_A$ (left plot/black, left plot/red, right plot respectively) to left-handed stop soft squared mass $m_{\tilde{Q}}^2$, right-handed stop soft squared mass $m_{\tilde{u}}^2$ and stop trilinear coupling $A_t$ as defined in Eq.\ \ref{eq:fQfUfA}. 
For the ellipse of low fine-tuning the rough form of Eq.~(\ref{eq:rough_ellipse}) was used and we have defined $\phi = \arg(\eta_1 + \text{i} \eta_2)$. 
The corresponding value of $f_x$ for $\eta_1 = \eta_2 = 1$ is shown as dashed line in the same color. Note that $f_u$ has been rescaled by a factor $5$ for illustration. 
}\label{fig:fQ_fU_fA}
\end{figure}

In the light of the discovery of a possible 125$\div$126 GeV Higgs boson in current LHC searches, it is interesting to re-analyse the consequences of non-universal gaugino masses for naturalness of the (non-universal) MSSM. In Fig.\ \ref{fig:AfterMh} we show the same plots as in Fig.\ \ref{fig:BeforeMh} with the additional experimental constraint from CMS, $m_h = 125.3 \pm 0.6 $ GeV \cite{discovery}, on top of which we include a theoretical uncertainty of $\pm 3$ GeV \cite{Arbey:2012dq} for the Higgs mass calculation at each data point.

As we can see from Fig.\ \ref{fig:AfterMh}a, non-universal gaugino masses can accommodate this additional constraint even with a fine-tuning lower than 20 (actually just above 10 and in a very small region of parameter space even slightly below 10).
This happens e.g., around $\eta_2 \sim 0$, $\eta_1 \sim 15$ and to a lesser degree also around $\eta_1 \sim 0$, $\eta_2 \sim 2.6$ and in the intermediate part of the ellipse. The reasons why these regions are favoured after including the Higgs results can readily be seen from the beta functions \cite{Martin:1993zk} of the stop soft masses masses and the stop trilinear coupling at the GUT scale:
\begin{subequations}\label{eq:fQfUfA}
\begin{align}
16 \pi^2 \beta_{m_{\tilde{Q}^2}} &\supset - g_{\text{GUT}}^2 M_3^2 \left(\frac{32}{3} + 6 \eta_2^2 + \frac{2}{15} \eta_1^2\right) \equiv - g_{\text{GUT}}^2 \, M_3^2 \, f_{Q}(\eta_1, \eta_2)\;, \\
16 \pi^2 \beta_{m_{\tilde{u}^2}} &\supset - g_{\text{GUT}}^2 M_3^2 \left(\frac{32}{3} + \frac{2}{15} \eta_1^2\right) \equiv - g_{\text{GUT}}^2 \, M_3^2 \, f_{u}(\eta_1, \eta_2) \;, \\
16 \pi^2 \beta_{A_t} &\supset - g_{\text{GUT}}^2 M_3 \left(\frac{32}{3} + 6 \eta_2 + \frac{26}{15} \eta_1\right) \equiv -g_{\text{GUT}}^2 M_3 \, f_{A}(\eta_1, \eta_2)\;. 
\end{align}
\end{subequations}
The functional behaviour of $f_Q$, $f_u$ and $f_A$ with $\eta_1 = r_\eta \cos\phi, \eta_2 = r_\eta \sin\phi$ and by means of the approximate form of the numerically obtained ellipse of Eq.~(\ref{eq:rough_ellipse}), is shown in Fig.\ \ref{fig:fQ_fU_fA}. 
It can be easily seen that the gaugino mass contribution to right-handed stop masses is greatly enhanced for the angles $\phi = 0, \pi$, i.e. $\eta_2 = 0$, $|\eta_1| \sim 15$, while it is not significantly away from the CMSSM value elsewhere. 
The contribution to the left-handed stop mass on the other hand receives a smaller but still sizable enhancement over the CMSSM value, but does not vary much along the ellipse. Additionally for $\phi = 0$, i.e. $\eta_1 \sim 15$ and $\eta_2 = 0$, the running contribution to $A_t$ receives a sizeable enhancement, which is present to a lesser degree also for $\phi = \pi / 2$, i.e. $\eta_1 = 0$ and $\eta_2 \sim 2.6$. For angles in the other quadrants, where at least one $\eta_i$ is negative, however, the cancellation with the large gluino mass running contribution prevents from a similar enhancement in $f_A$. However, in practice, these three effects turn out to be effective only when they work in unison to obtain a high enough Higgs mass.

\begin{figure}
\centering
\includegraphics[width=0.6\textwidth]{plots/NUGM-legend.pdf}

\includegraphics[width=0.47\textwidth]{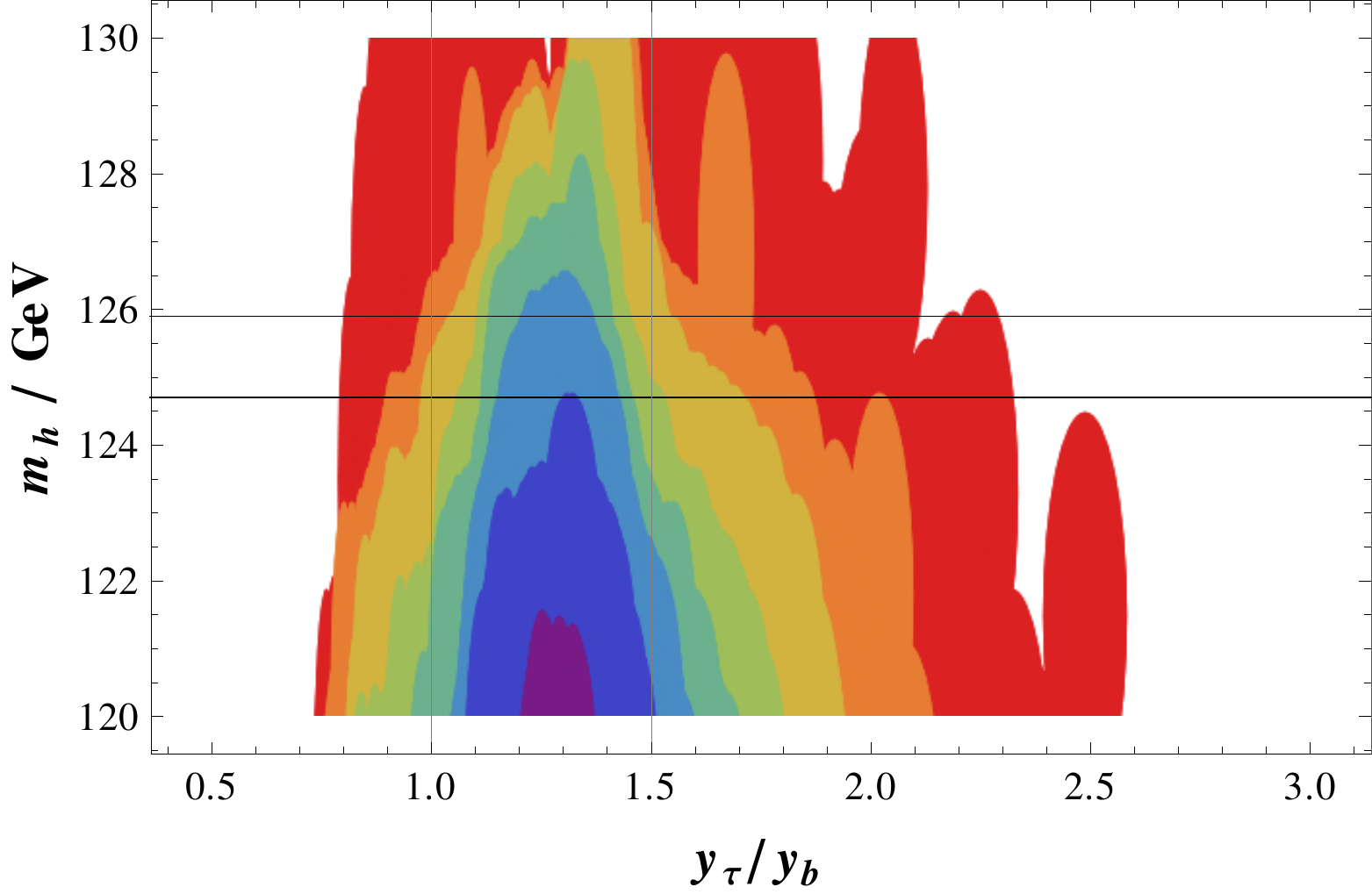}
\includegraphics[width=0.47\textwidth]{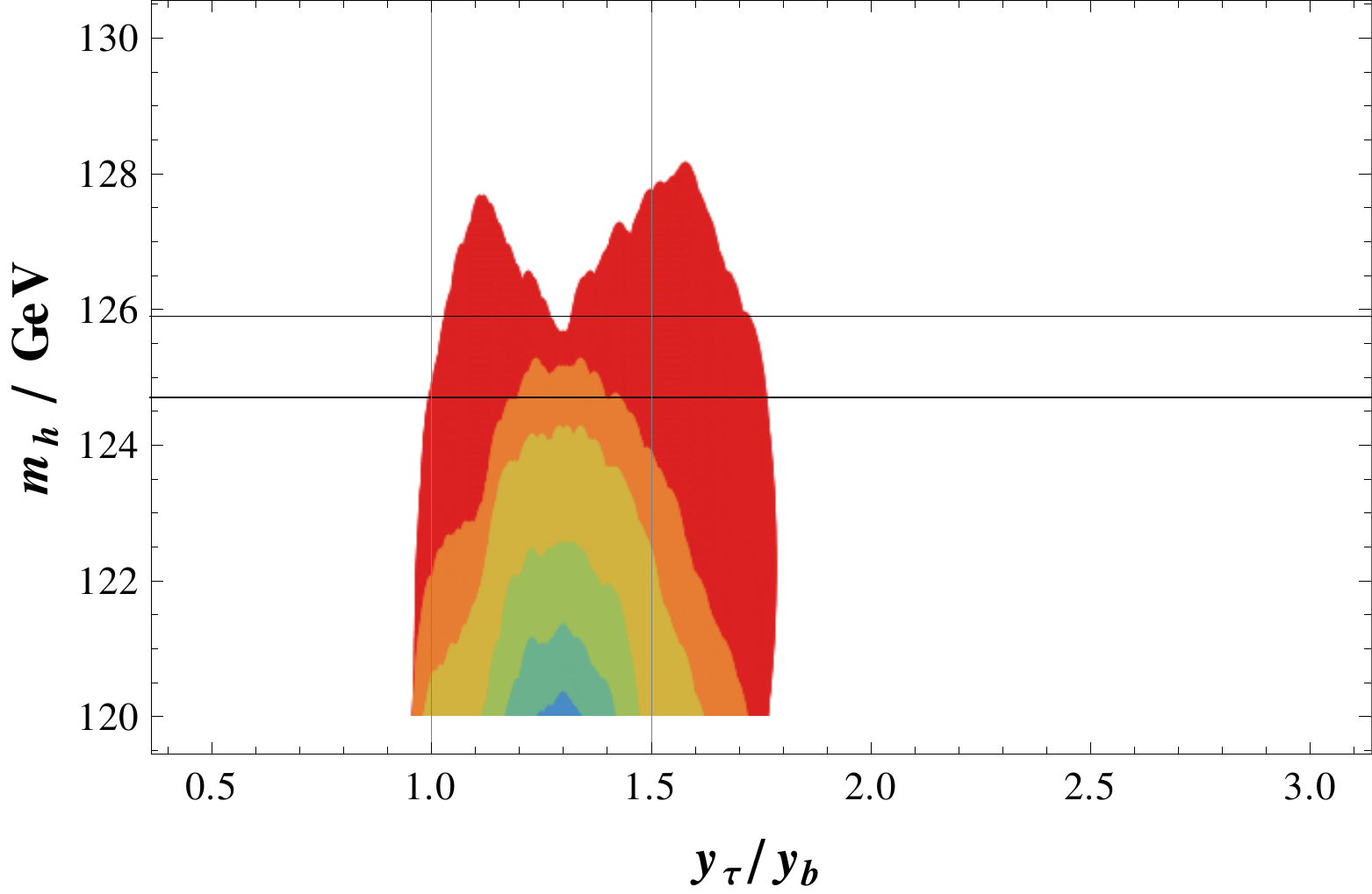}
\caption{
Lowest fine-tuning in the $m_{h}$-$y_\tau / y_b$ plane consistent 
with experimental bounds described in the text for non-universal gaugino masses (left) and universal gaugino masses (right). The horizontal lines correspond to the 1$\sigma$ error of $0.6$ GeV around $125.3$ GeV as claimed by the CMS collaboration\cite{discovery}.
The 1$\sigma$ errors on the quark masses \cite{PDG} are taken into account
by scaling the data points correspondingly. In addition, a theoretical uncertainty of $3$ GeV \cite{Arbey:2012dq} for the Higgs mass calculation at each data point is taken into account.
}\label{fig:mh0overytauyb}
\end{figure}

We note that the changes in Fig.\ \ref{fig:AfterMh}b are not very suprising: regions with low masses get cut off. Also the changes in Fig.\ \ref{fig:AfterMh}c are not unexpected, the solution with $\eta_A > 0$ is disfavoured more than the solution with $\eta_A \sim 0$, as the latter does not suffer from the cancellation with the gluino mass contribution to the running of the top trilinear soft term. Unsurprisingly, it is shown that either large $\eta_A$ or large $M_3$ (hence large top trilinear coupling) is needed to accommodate a large Higgs mass.

Concerning the Yukawa coupling ratios shown in Fig.\ \ref{fig:AfterMh}d we can see that the minimal fine-tuning required to get to a unified $b$-$\tau$ Yukawa coupling ratio suffers more from the requirement of a consistent Higgs mass than the ratio of $3/2$ does. Namely after applying the cut the fine-tuning for $y_\tau / y_b = 3/2$ can go down to $\Delta = 30$, while $y_\tau = y_b$ requires at least a $\Delta$ of $60$. This is also nicely illustrated in Fig.\ \ref{fig:mh0overytauyb} where we have shown the interplay among the Higgs mass, the GUT scale $y_\tau/y_b$ ratio and the amount of fine-tuning.

\begin{figure}
\vspace{-1.8cm}
\centering
\quad \quad \quad \:\includegraphics[width=0.6\textwidth]{plots/NUGM-legend.pdf}

\begin{tabular}{llll}
\begin{sideways}\quad\quad\quad\footnotesize w/o Higgs Results\end{sideways} &
\includegraphics[width=0.3\textwidth]{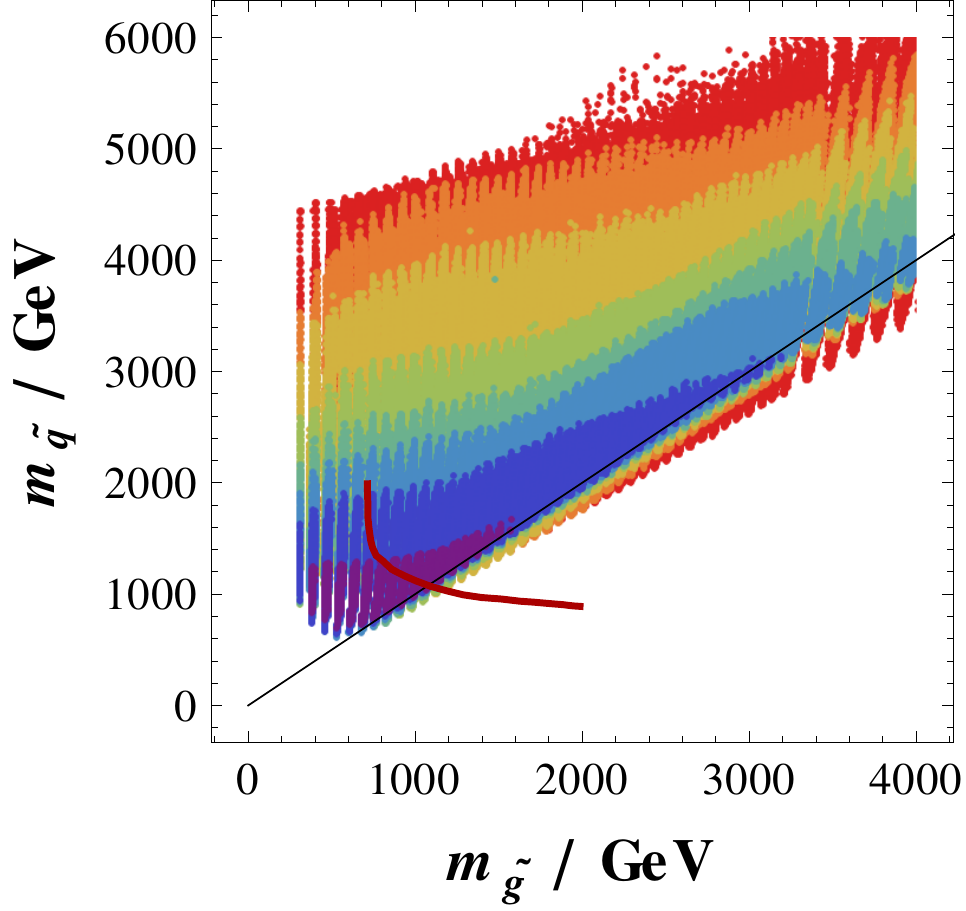} &
\includegraphics[width=0.3\textwidth]{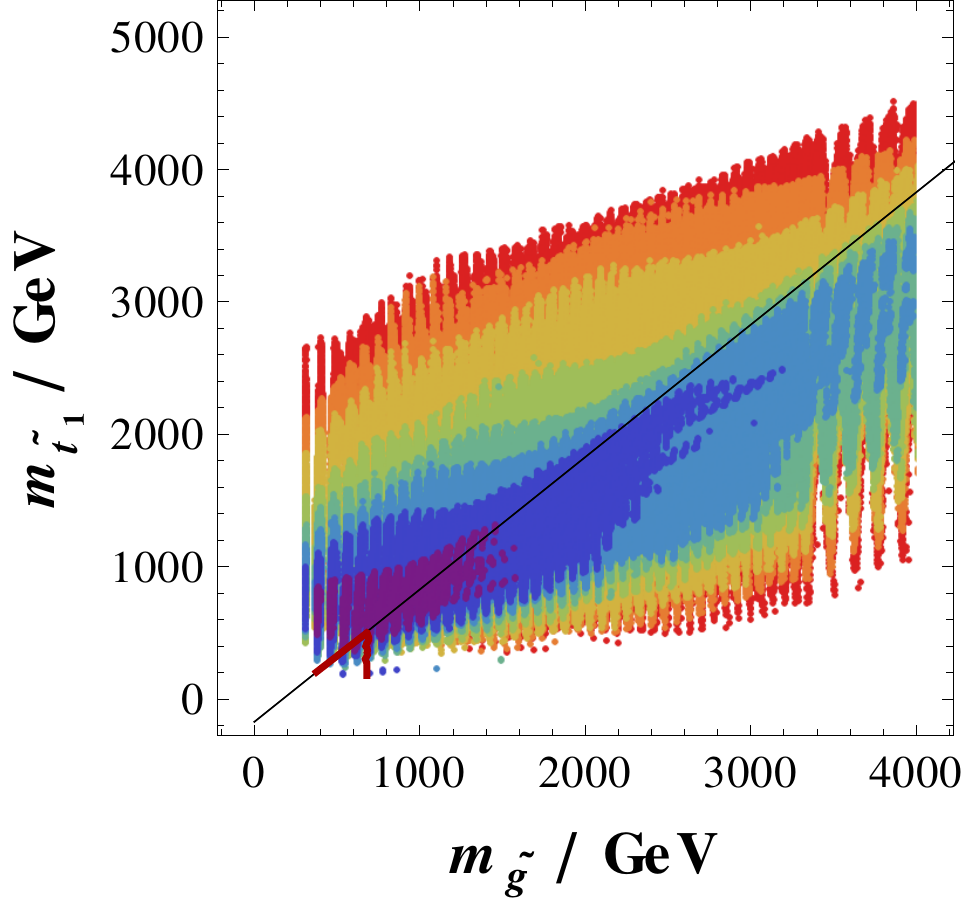} &
\includegraphics[width=0.3\textwidth]{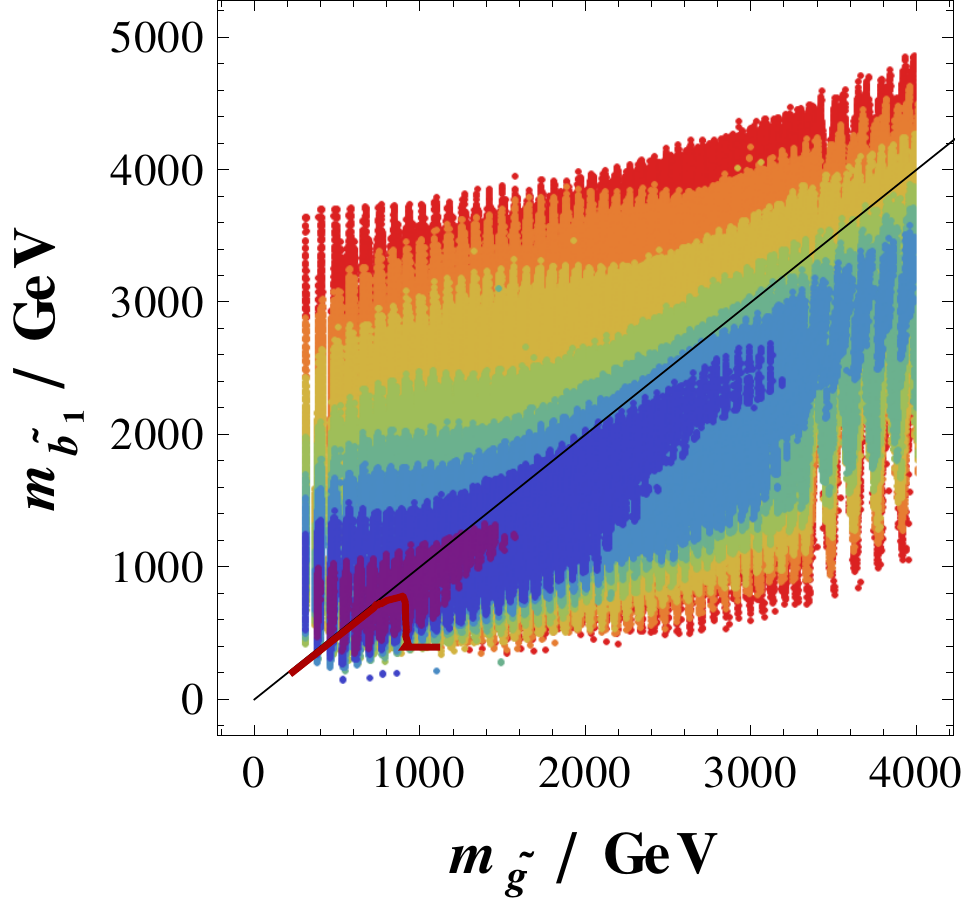} \\

\begin{sideways}\enspace\,\quad \:\quad \footnotesize  with Higgs Results\end{sideways} &
\includegraphics[width=0.3\textwidth]{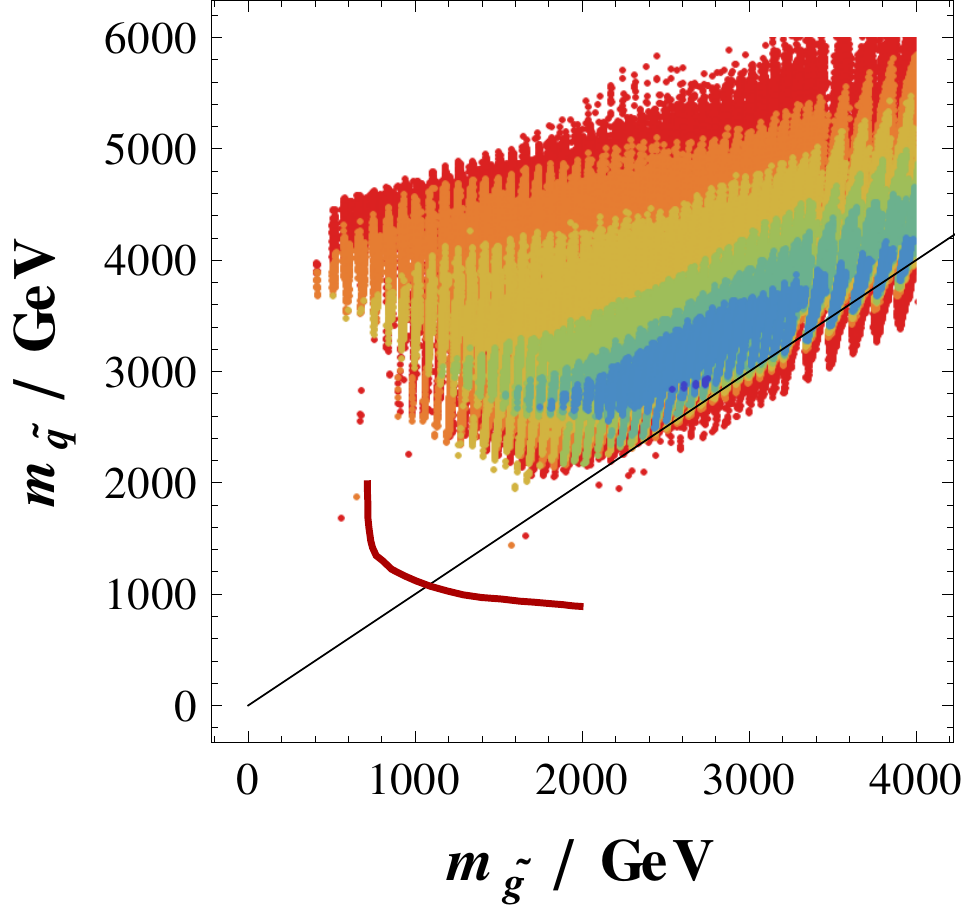} &
\includegraphics[width=0.3\textwidth]{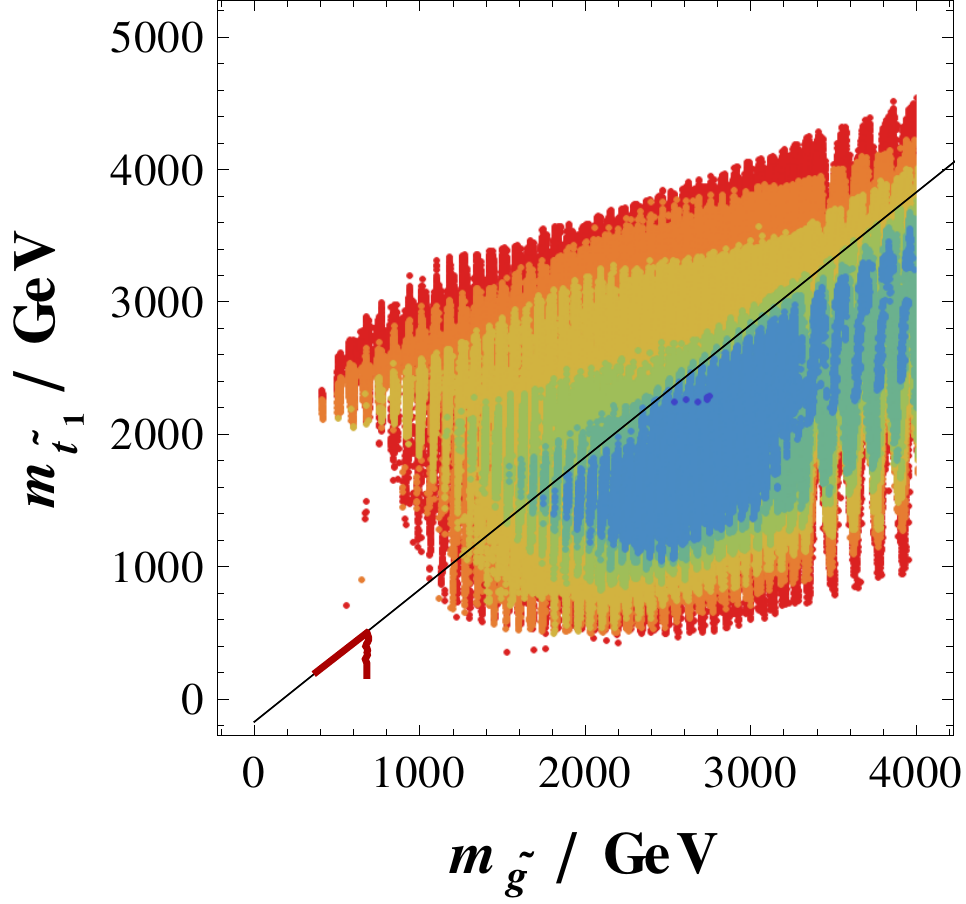} &
\includegraphics[width=0.3\textwidth]{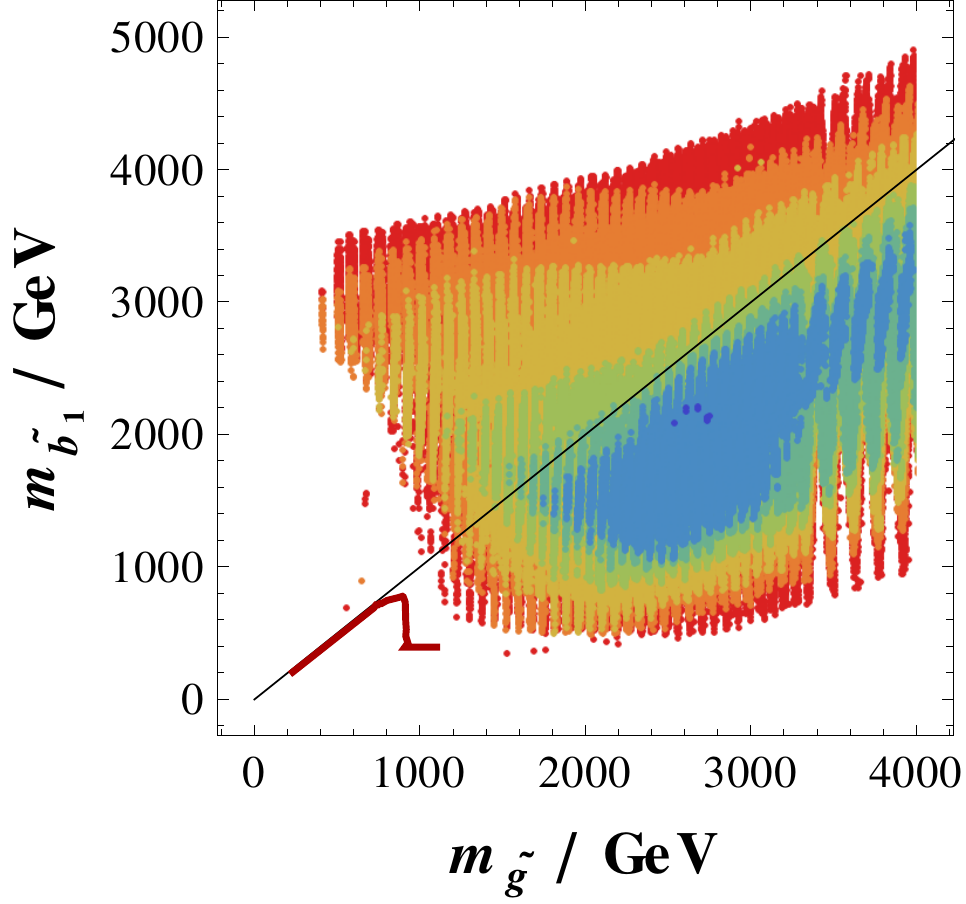}
\end{tabular}

\begin{tabular}{lll}
\begin{sideways}\quad\quad \quad \footnotesize w/o Higgs Results\end{sideways}&
\includegraphics[width=0.3\textwidth]{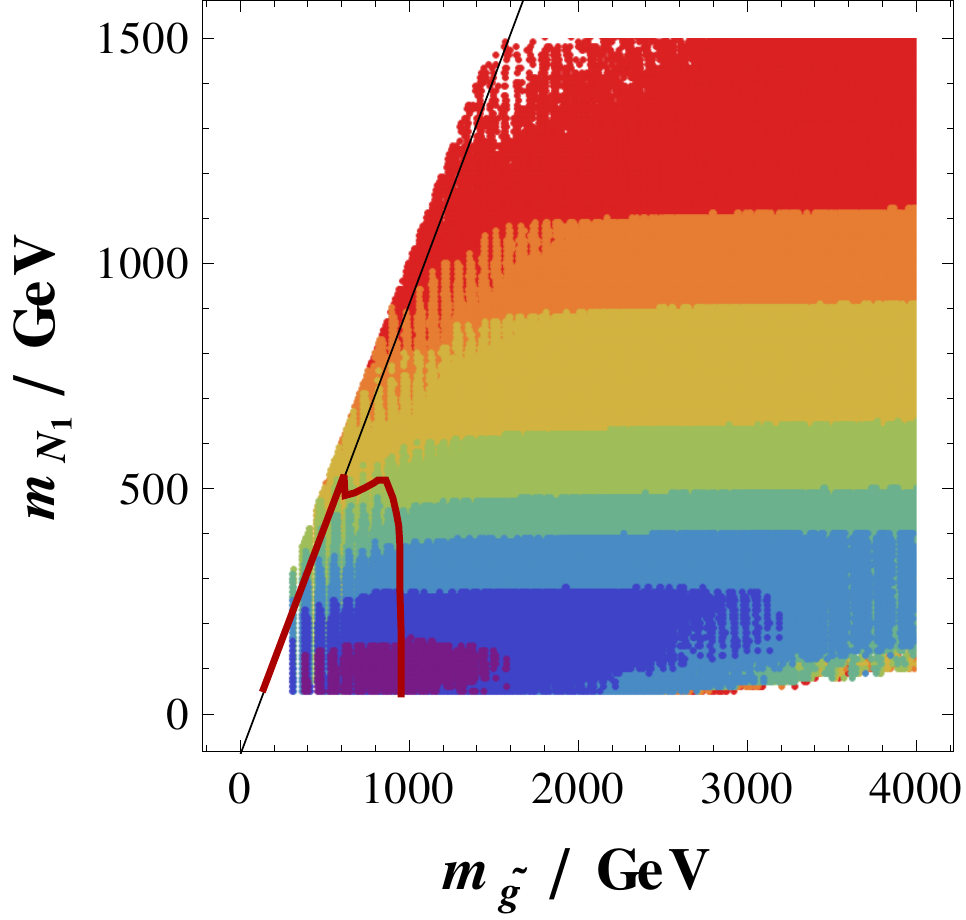} &
\includegraphics[width=0.3\textwidth]{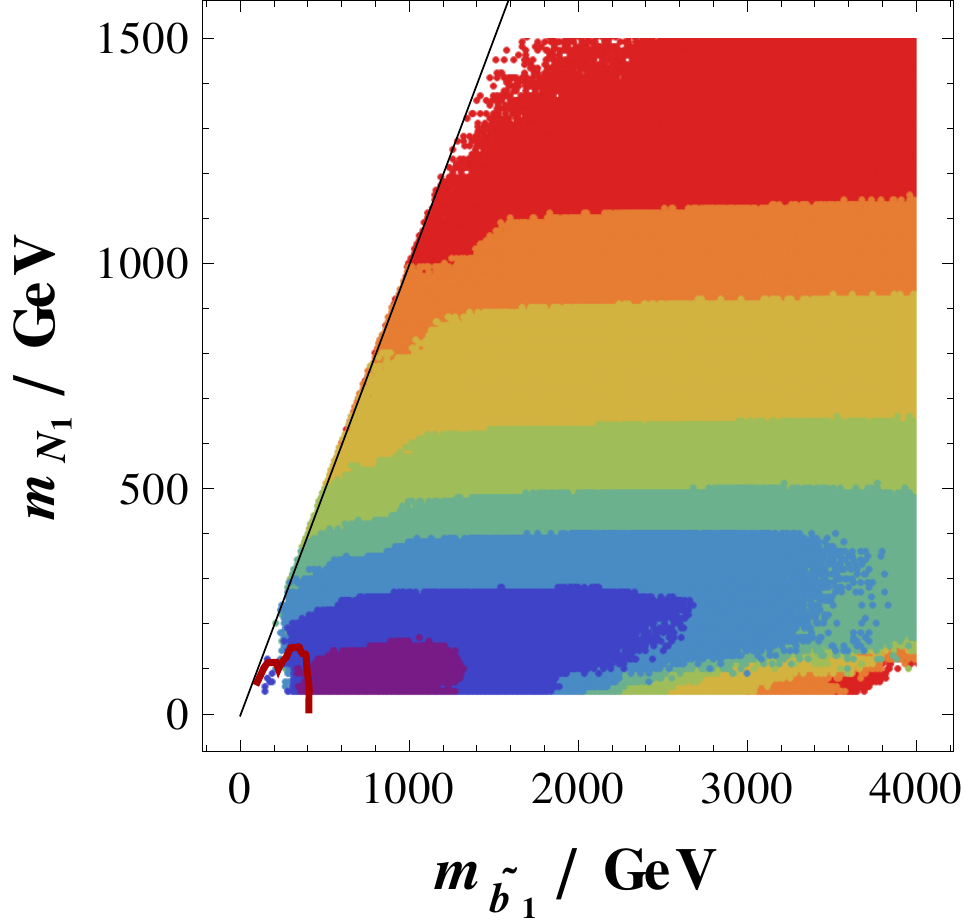} \\

\begin{sideways}\enspace\,\quad \:\quad \footnotesize with Higgs Results\end{sideways}&
\includegraphics[width=0.3\textwidth]{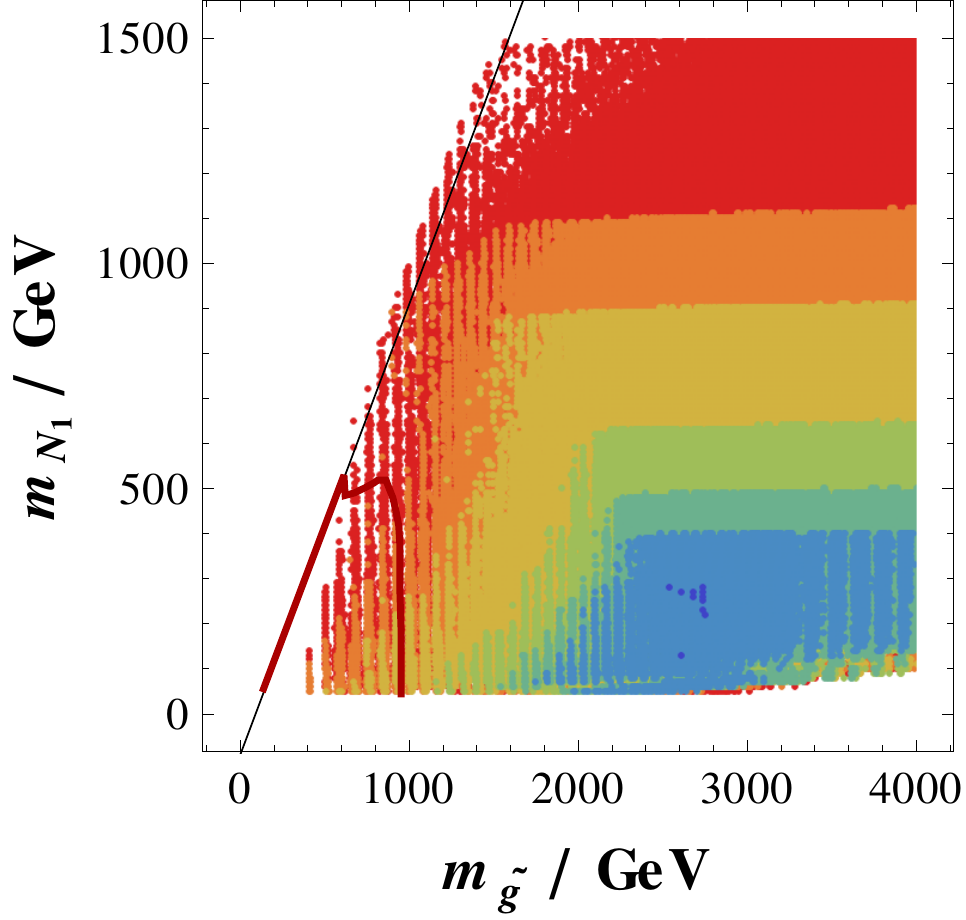} &
\includegraphics[width=0.3\textwidth]{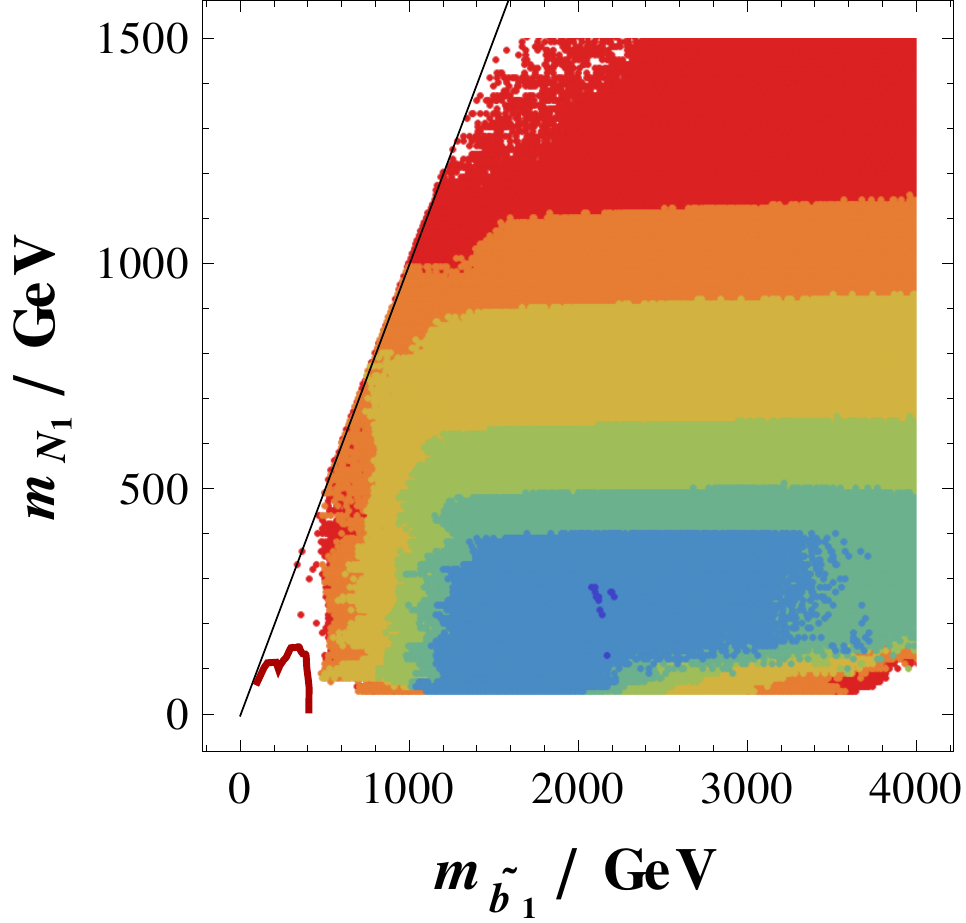}
\end{tabular}
\caption{
Lowest fine-tuning in various planes used for simplified models. Only points consistent with experimental bounds described in the text are shown.
In addition, for the plots on the second and fourth line we have included the experimental constraint $m_h = 125.3 \pm 0.6 $ GeV \cite{discovery} and a theoretical uncertainty of $\pm 3$ GeV \cite{Arbey:2012dq} for the Higgs mass calculation at each data point. 
The corresponding bounds due to LHC SUSY searches are shown as thick dark red lines \cite{SMSbounds}.
Note that the bounds can only be used to give a general flavour on how the direct SUSY searches at the LHC affect our results because we did not do a full detector simulation. For more details see main text.
}\label{fig:simplifiedmodelbounds}
\end{figure}

Another important aspect of current LHC experiments are the searches for supersymmetric particles.
Since changes of the gaugino mass ratios $\eta_1$ and $\eta_2$ can significantly alter the composition of the lightest neutralino\footnote{Actually, low fine-tuning generally also means quite low $\mu$, so Higgsino-like lightest neutralinos and charginos are likely.} as well as the mass splittings controlling jet energies and missing $E_T$ from the cascade decays, the exclusion regions for the CMSSM do not apply anymore. A full event and detector simulation would, however, go beyond the scope of this study; thus we make use of exclusion bounds derived in so-called simplified models \cite{SMS,SMSbounds}. 
For simplicity we just compare the spectra found by the numerical scan with the most stringent bounds in several kinds of simplified models. 
While this is not certainly a rigorous approach, 
it should give a good feeling how endangered by exclusion each point is. 

The resulting situation is shown in Fig.\ \ref{fig:simplifiedmodelbounds}. 
As we can see in most cases only the region with $\Delta < 3$ is partially inside the excluded region, while even $\Delta < 10$ extends far beyond the bounds. Requiring a Higgs boson within 1$\sigma$ of the experimental measurement \cite{discovery} (including 3 GeV theoretical uncertainty as explained before) excludes even more than what direct SUSY searches do in some areas. It is thus important to note that, even in this quite restrictive approach, all of the parameter space with high $m_h$ and $\Delta < 20$ is safe from being excluded by the LHC in the near future.

\begin{figure}
\centering
\includegraphics[width=0.6\textwidth]{plots/NUGM-legend.pdf}

\includegraphics[width=0.47\textwidth]{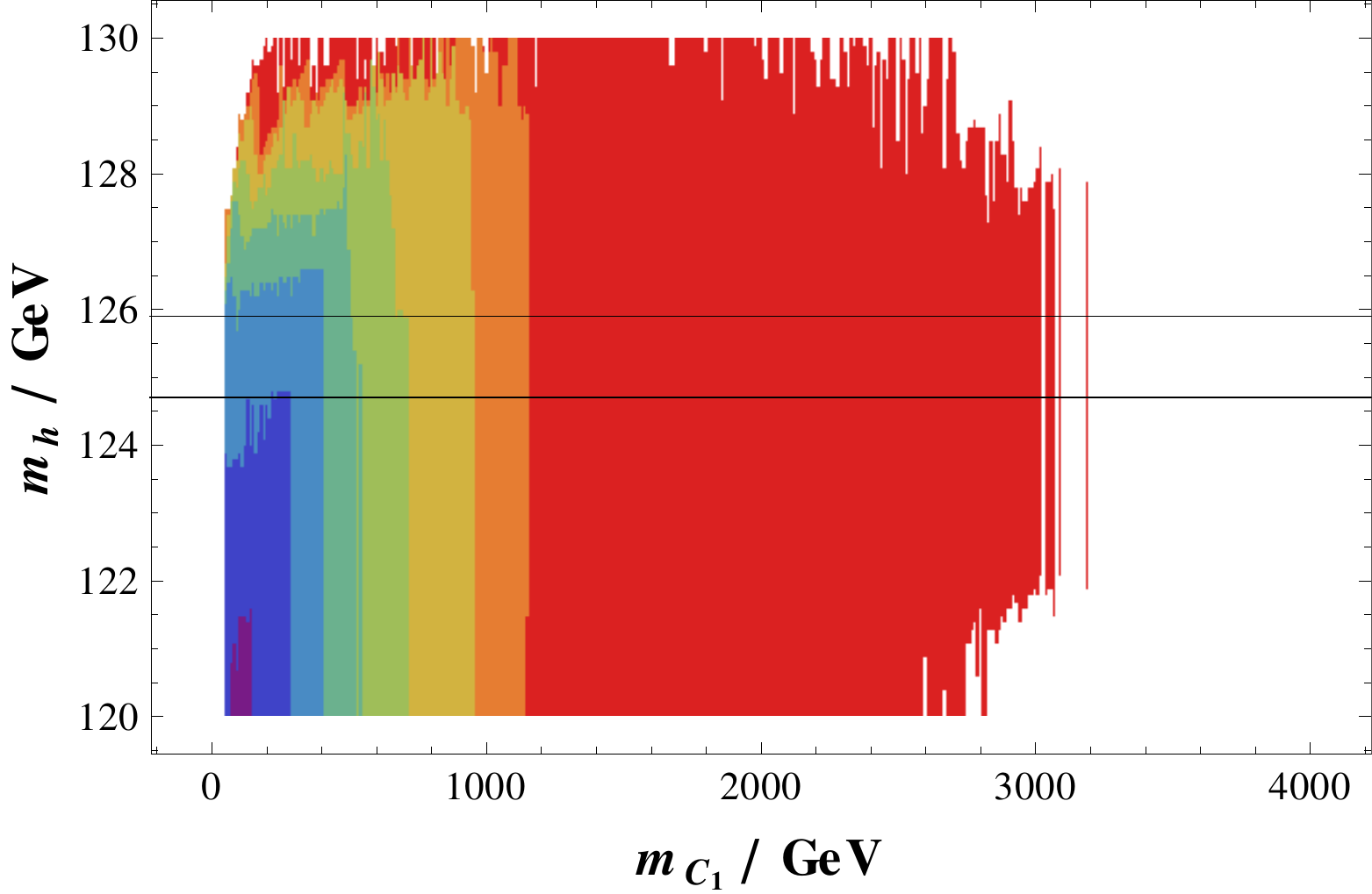}
\includegraphics[width=0.47\textwidth]{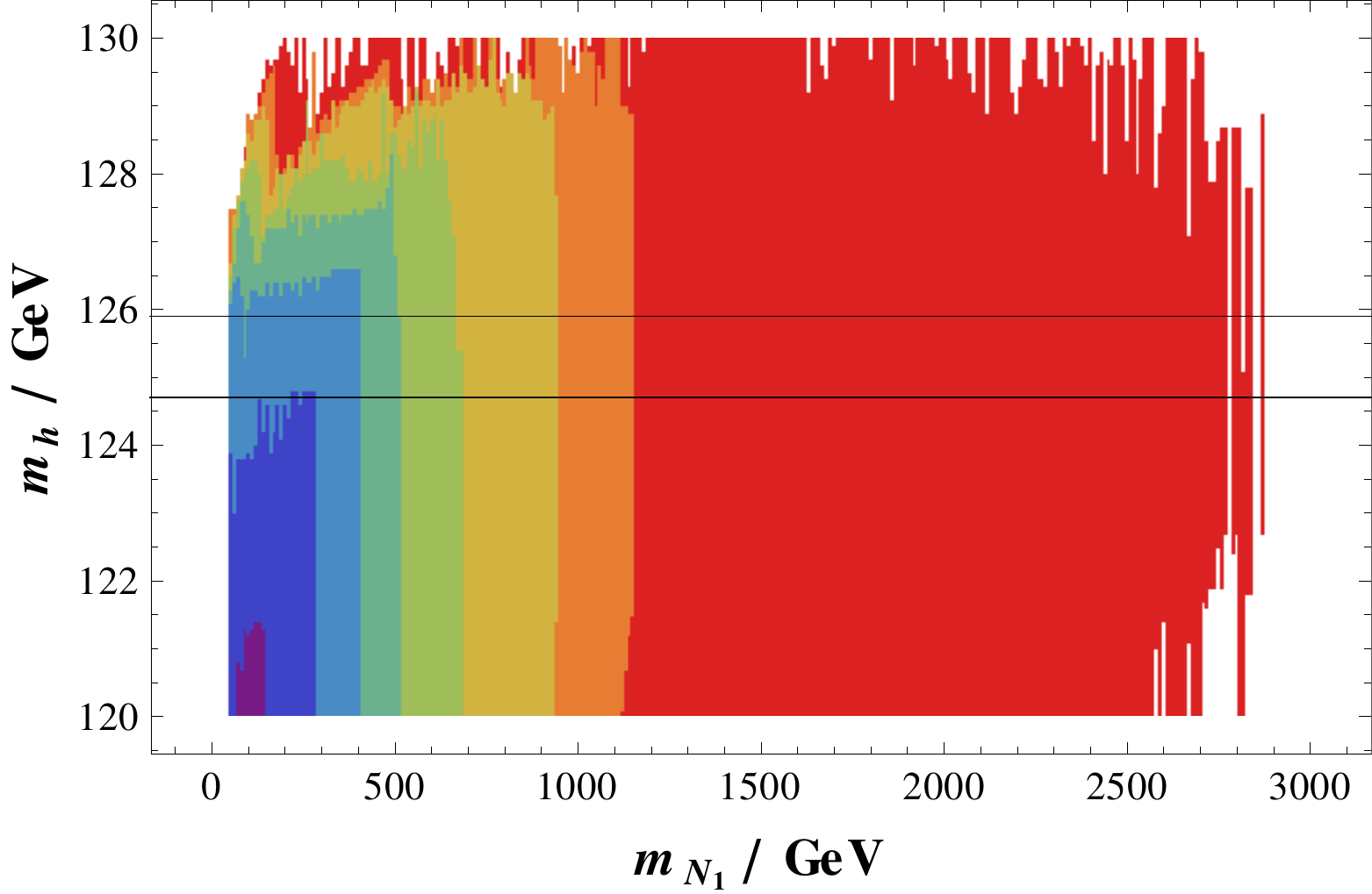}

\includegraphics[width=0.47\textwidth]{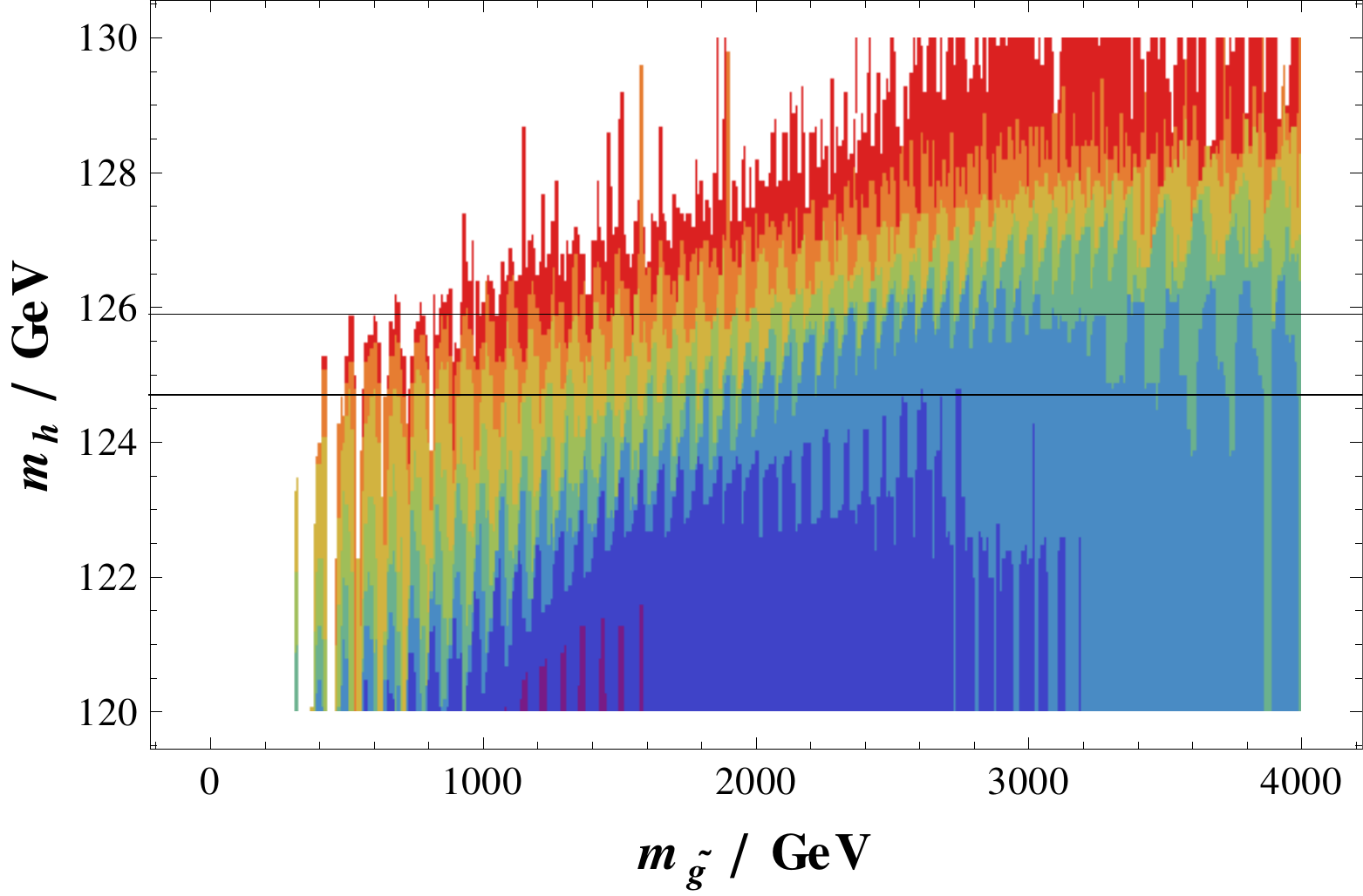}
\includegraphics[width=0.47\textwidth]{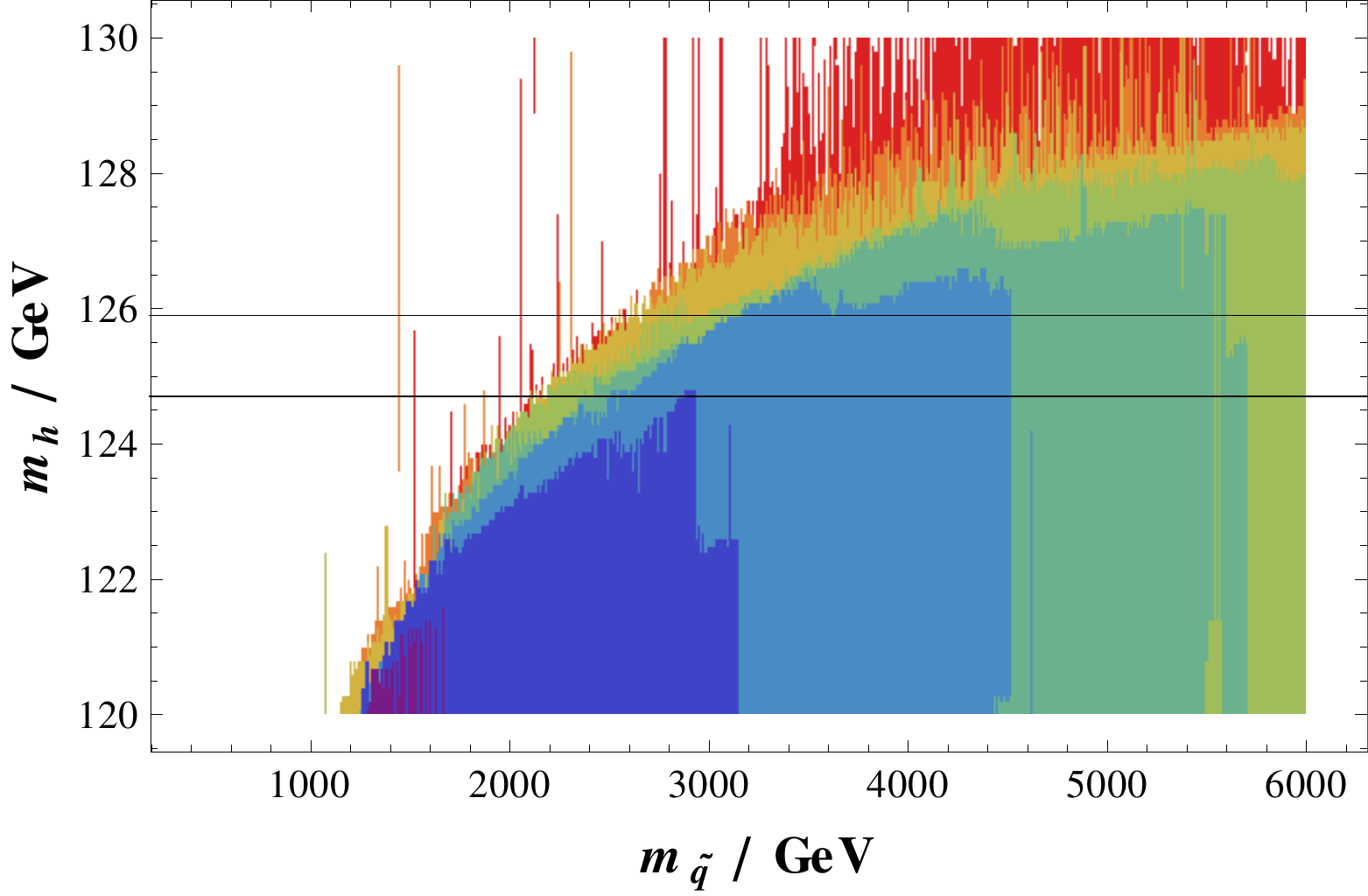}

\includegraphics[width=0.47\textwidth]{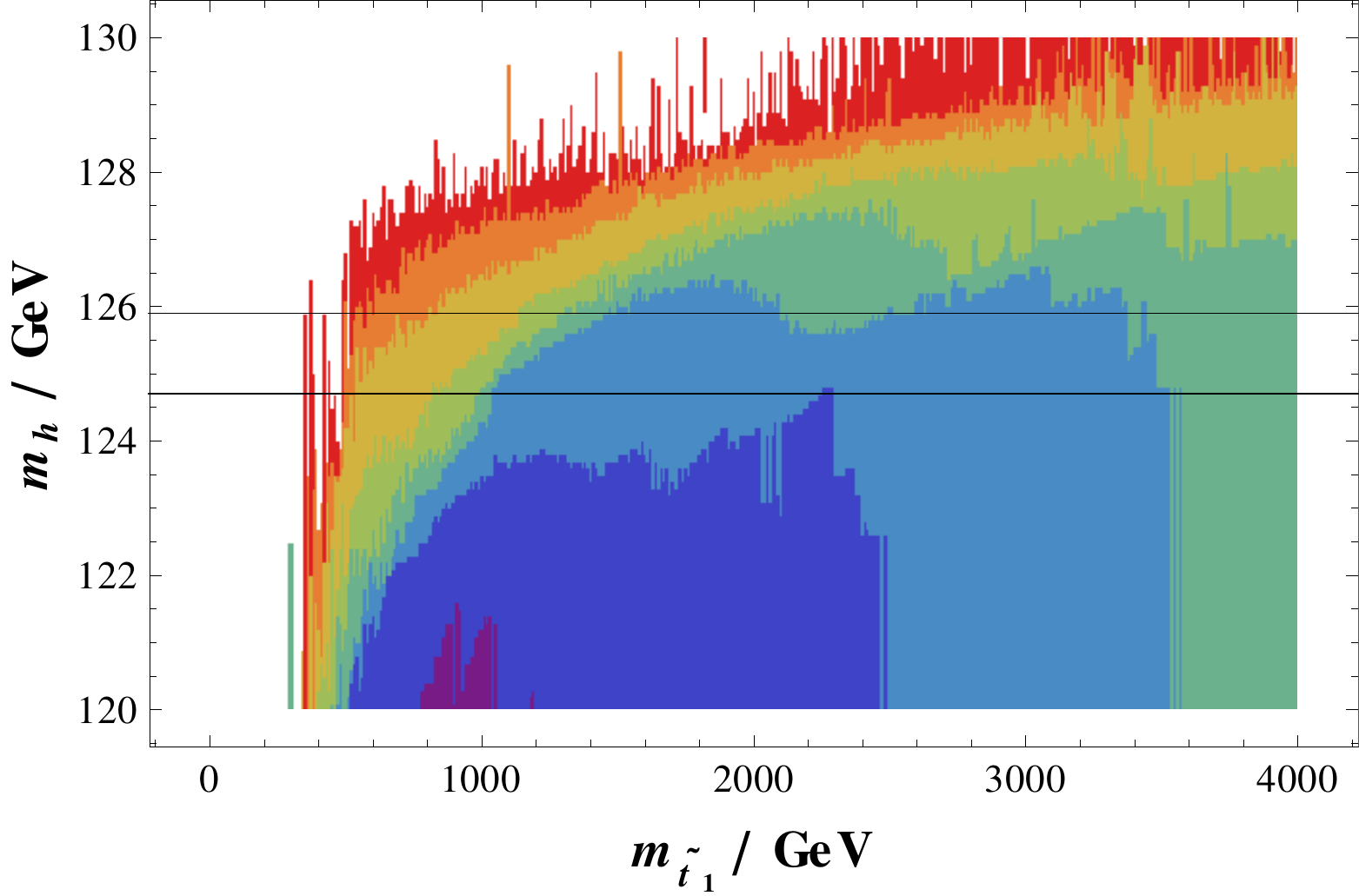}
\includegraphics[width=0.47\textwidth]{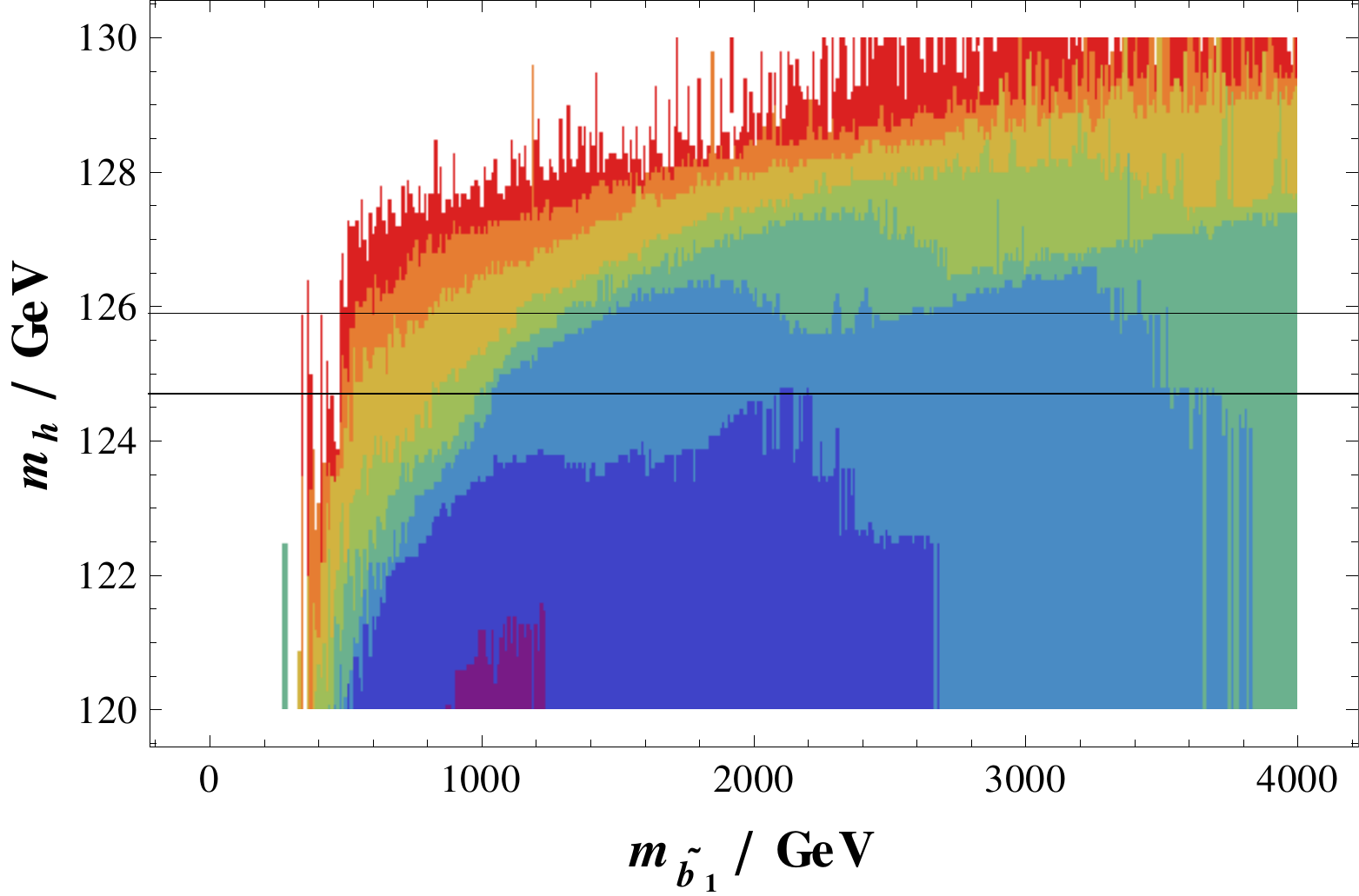}
\caption{
Lowest fine-tuning shown for the Higgs mass vs.~important sparticle masses including the experimental constraint $m_h = 125.3 \pm 0.6 $ GeV \cite{discovery} and a theoretical uncertainty of $\pm 3$ GeV \cite{Arbey:2012dq} for the Higgs mass calculation at each data point.
}\label{fig:higgs_vs_othermass}
\end{figure}

For future searches it is also interesting to look at the correlation between the Higgs and several sparticle masses, which is shown in Fig.\ \ref{fig:higgs_vs_othermass}. There we can see that for low fine-tuning relatively light neutralinos and charginos are expected. This is due to the fact that $\mu$ is small in this region, so that we expect Higgsinos 
to be light. We also see that the gluinos and the lightest stop are rather heavy in our scenario with low fine-tuning ($\Delta < 20$ for $m_{\tilde{g}} \gtrsim 1.5$~TeV and $m_{\tilde{t}_1} \gtrsim 1.0$~TeV). The LHC still did not reach this parameter region and hence the statement that the natural MSSM parameter space is already ruled out seems to be premature.

\subsection{Favoured Non-Universal Gaugino Mass Ratios}

\begin{table}
\begin{center}
\begin{tabular}{ccl}
\toprule
$\eta_1$, $\eta_2$ & $\Delta_{\min}$ & Origin \\
\midrule
1, 1 & 118 & CMSSM (Gaugino Unification) \\
\midrule
10, 2 & 12 & 200 of SU(5) \quad \cite{gauginoratios}\\[0.5ex]
$\frac{19}{10}$, $\frac{5}{2}$ & 18 & 770 of $SO(10) \to (1, 1)$ of $SU(4) \times SU(2)_R$ \quad \cite{gauginoratios}\\[0.5ex]
$\frac{77}{5}$, 1 & 36 & 770 of $SO(10) \to (1, 0)$ of $(SU(5)'  \times  U(1))_{\text{flipped}}$\quad \cite{gauginoratios}\\[0.5ex]
$-\frac{1}{5}$, 3 & 46 & 210 of $SO(10) \to (75, 0)$ of $(SU(5)'  \times U(1))_{\text{flipped}}$\quad \cite{gauginoratios}\\
\midrule
$\frac{21}{5}$, $\frac{7}{3}$ & 13 & O-II with $\delta_{\text{GS}} = -6$ \quad \cite{Brignole:1993dj,Horton:2009ed}\\[0.5ex]
$\frac{17}{5}$, $2$ & 28 & O-II with $\delta_{\text{GS}} = -7$ \quad \cite{Brignole:1993dj,Horton:2009ed}\\[0.5ex]
$\frac{29}{5}$, $3$ & 44 & O-II with $\delta_{\text{GS}} = -5$ \quad \cite{Brignole:1993dj,Horton:2009ed}\\
\bottomrule
\end{tabular}
\end{center}
\caption{
Selected ratios and the minimal possible fine-tuning they allow after requiring the experimental constraint 
$m_h = 125.3 \pm 0.6 $ GeV \cite{discovery} and a theoretical uncertainty of $\pm 3$ GeV \cite{Arbey:2012dq} 
for the Higgs mass calculation.
Only ratios that can reduce the fine-tuning by at least 50\% compared to the unified (CMSSM) scenario are shown. 
For more details on the origin of these ratios, see, e.g.\ \cite{gauginoratios,Horton:2009ed,Brignole:1993dj}. The results are illustrated graphically in Fig.\ \ref{fig:eta2overeta1_comparison}.
}\label{tab:ratiocomparison}
\end{table}

\begin{figure}
\centering
\includegraphics[width=0.6\textwidth]{plots/NUGM-legend.pdf}

\includegraphics[width=0.75\textwidth]{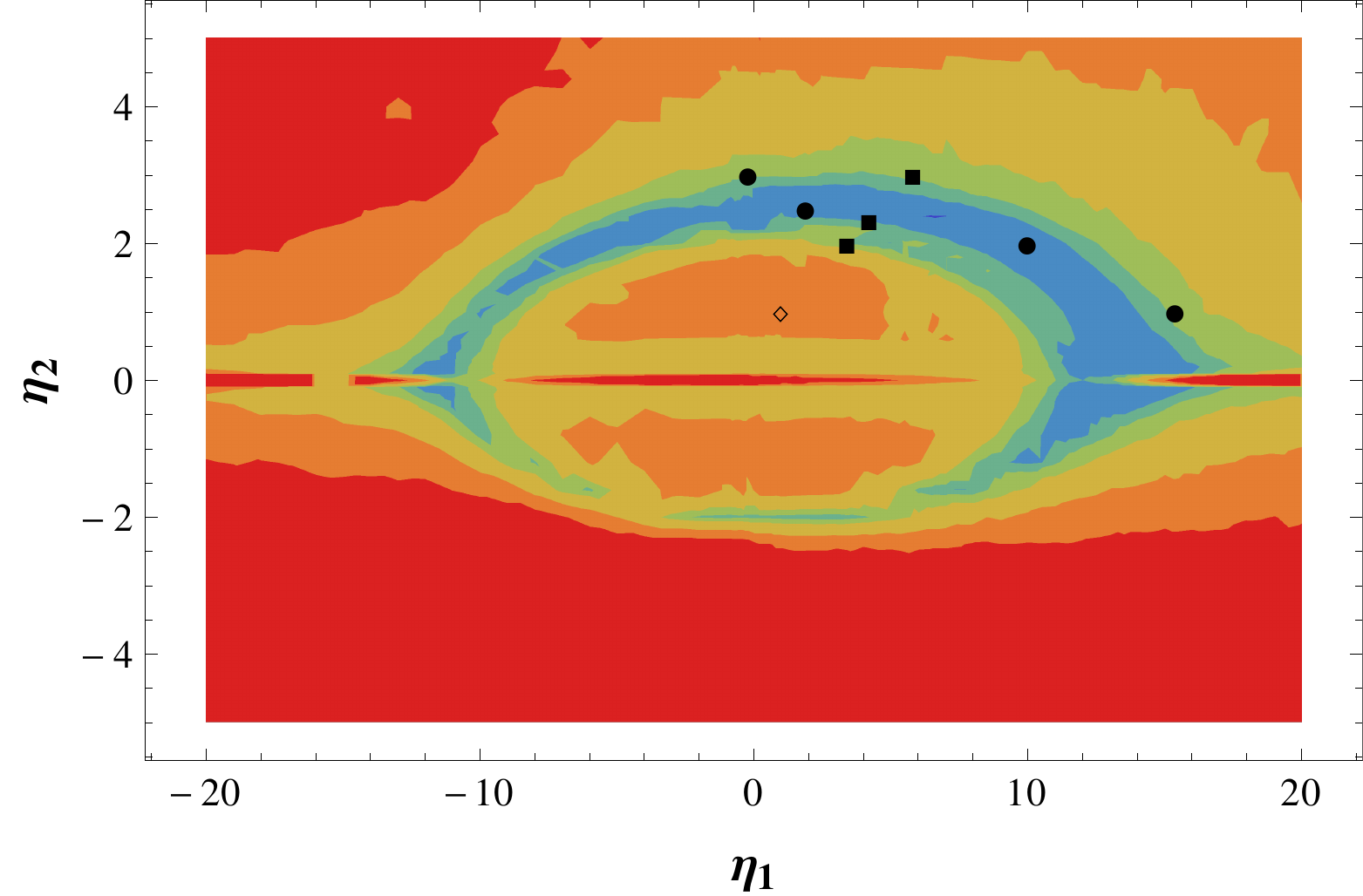} \quad \quad \quad \quad
\caption{
Lowest fine-tuning in the $\eta_1$-$\eta_2$ plane consistent 
the experimental constraint $m_h = 125.3 \pm 0.6 $ GeV \cite{discovery} and a theoretical uncertainty of $\pm 3$ GeV \cite{Arbey:2012dq} 
for the Higgs mass calculation at each data point. 
Theoretically motivated ratios that reduce the fine-tuning compared to the unified scenario by at 
least 50\% and the analytical expectation are also included.
The CMSSM ratio is marked as empty diamond, circles correspond to ratios derived from GUT 
symmetry breaking \cite{gauginoratios} and squares to ratios found in the so called O-II model in \cite{Brignole:1993dj,Horton:2009ed}. 
For details see Tab.\ \ref{tab:ratiocomparison}.
}\label{fig:eta2overeta1_comparison}
\end{figure}

Fixed non-universal gaugino mass ratios may originate from various high energy models, for instance from GUTs or orbifold scenarios (see, e.g.\ \cite{gauginoratios,Horton:2009ed,Brignole:1993dj} for discussions). Tab.\ \ref{tab:ratiocomparison} and Fig.\ \ref{fig:eta2overeta1_comparison} show examples of proposed fixed ratios $\eta_1,\eta_2$ where we find that the fine-tuning can be reduced by more than 50\% compared to the CMSSM. 

Interestingly, among the orbifold models O-II of \cite{Brignole:1993dj,Horton:2009ed}, from the full numerical results including the constraints from the latest Higgs results we find that the option $(\eta_1,\eta_2) = (\tfrac{21}{5},\tfrac{7}{3})$ from $\delta_{\text{GS}} = -6$ has the lowest possible fine-tuning ($\Delta_{\text{min}} = 13$), whereas in \cite{Horton:2009ed}, based on analytical estimates before the Higgs results were available, the preferred ratio was $(\eta_1,\eta_2) = (\tfrac{29}{5},3)$ from $\delta_{\text{GS}} = -5$.\footnote{$\delta_{\text{GS}}$ is a negative integer constant associated with Green-Schwarz anomaly cancellation (cf.\ \cite{Brignole:1993dj,Horton:2009ed}).}

The GUT ratio with lowest fine tuning ($\Delta_{\text{min}} = 12$) turned out to be $(\eta_1,\eta_2) = (10,2)$. For the ratios found to be favoured in \cite{Horton:2009ed}, we find significantly higher fine-tuning, e.g.: 
$\Delta_{\text{min}} = 82$ for $(\eta_1,\eta_2) = (-5,3)$, 
$\Delta_{\text{min}} = 141$ for $(\eta_1,\eta_2) = (-\tfrac{101}{10},-\tfrac{3}{2})$, 
$\Delta_{\text{min}} = 143$ for $(\eta_1,\eta_2) = (1,-\tfrac{7}{3})$.

As it can be seen as well from comparison of Fig.~\ref{fig:BeforeMh}a and \ref{fig:AfterMh}a, the inclusion of the Higgs results has strong effects on the favoured non-universal gaugino mass ratios.

Finally, as discussed at the beginning of section \ref{sec:num}, for $\eta_2 < 0.5\, \eta_1$ the neutralino is dominated by its Wino (or Higgsino) component which implies that the relic density is strongly suppressed and there is therefore no danger of overproducing thermal neutralino dark matter. Among the ratios listed in Tab.\ \ref{tab:ratiocomparison} this applies to $(\eta_1,\eta_2) = (10,2)$ and $(\eta_1,\eta_2) = (\frac{77}{5},1)$. For the other ratios of Tab.\ \ref{tab:ratiocomparison}, $\eta_2 > 0.5\, \eta_1$ holds and, at least in principle, thermal neutralino dark matter could be possible (for a standard thermal history of the universe). As already mentioned, we have not applied dark matter constraints in our analysis since they depend on the thermal history of the universe.

\section{Summary and Conclusions}

We have re-addressed the question of the naturalness of the MSSM,
in the light of the discovery of a Higgs-like resonance recently announced
by the LHC experiments \cite{discovery}. We focused on models with
non-universal boundary conditions of the SUSY breaking parameters at high
energy and compared them with the CMSSM. Our basic assumption was that the
departures from universality are a consequence of an underlying mechanism,
possibly associated to SUSY breaking or GUT dynamics, giving fixed relations
among the parameters. In order to identify which of such relations can lead
to a reduced fine-tuning, we first considered a general high-energy
parametrization of the soft terms. Then we discussed scenarios with
non-universal scalar and gaugino masses. We found this latter possibility
particularly promising (as already noticed, e.g., in \cite{Abe:2007kf, Gogoladze:2009bd, Horton:2009ed})
and we thus studied it in detail, computing numerically the fine-tuning
measure $\Delta$.

We found that, considering the uncertainty related to the theoretical
prediction, models with non-universal gaugino masses can account for a Higgs
mass in the CMS 1$\sigma$ range $125.3\pm 0.6$ with a fine-tuning price 
of ${\cal O}(10)$, in contrast to the CMSSM that requires $\Delta_{\text{min}} \gtrsim {\cal O}(100)$. Thus
the MSSM with specific gaugino mass ratios still represents a comparatively
natural scenario. In Fig.\ \ref{fig:BeforeMh} and \ref{fig:AfterMh}, we have shown the
values of $\eta_1 = M_1/M_3$ and $\eta_2= M_2/M_3$ giving a low fine-tuning
before and after applying a constraint on the Higgs mass.
Interestingly, some of the ratios discussed in the literature
lie in (or are close to) the new low fine-tuning region, while others are disfavoured when the new Higgs results are included. 
Including the Higgs results we found that particularly favoured ratios (with $\Delta_{\text{min}} = {\cal O}(10)$) 
are now, e.g., $(\eta_1,\ \eta_2) = (10, 2)$ 
which may originate from $SU(5)$ GUTs and $(\eta_1,\eta_2) = (\tfrac{21}{5},\tfrac{7}{3})$ 
from orbifold scenarios of type O-II with $\delta_{\text{GS}} = -6$. 

Furthermore, allowing for non-universal gaugino masses at the GUT scale, 
we have analyzed the fine-tuning price of different values of the
GUT-scale ratio $y_\tau/y_b$, that represents an important handle to
discriminate among different GUT models. For $m_h \approx 125$ GeV, we
found that $b$-$\tau$ Yukawa unification corresponds to $\Delta \gtrsim 60$,
while the alternative ratio $y_\tau/y_b= 3/2$ can be realized at the price
of $\Delta \gtrsim 30$ only.

Concerning the SUSY spectrum that is favoured by naturalness and $m_h \approx 125$ GeV,
we found that for the least tuned data points with fine-tuning $\Delta$ less than 20 the lightest neutralino 
is expected to be lighter than about 400 GeV and the lighter stop can be as heavy 
as 3.5 TeV. On the other hand, the gluino mass is required to be above 1.5 TeV. 
Comparing the predicted spectra with the LHC exclusions derived in a set of simplified models, we could conclude
that the regions of lowest fine-tuning are at present only poorly constrained by direct SUSY searches at the LHC.

Let us finally remark that, although the CMSSM is certainly challenged
from the fine-tuning point of view by
a Higgs mass of $m_h \approx 125$ GeV, more general realizations of the MSSM,
like the examples studied here, can still provide a relatively natural
solution of the hierarchy problem and will probably require several years
of data taking before being fully tested at the LHC.

\section*{Acknowledgements}

S.A.\ and V.M.\ were supported by the Swiss National Science Foundation.
S.A.\ and M.S.\ acknowledge partial support from the European Union
under FP7 ITN INVISIBLES (Marie Curie Actions,
PITN-GA-2011-289442).


\begin{thebibliography}{99}

\bibitem{discovery}
  G.~Aad {\it et al.}  [ATLAS Collaboration],
  Phys.\ Lett.\ B {\bf 716} (2012) 1
  [arXiv:1207.7214 [hep-ex]];
  S.~Chatrchyan {\it et al.}  [CMS Collaboration],
  Phys.\ Lett.\ B {\bf 716} (2012) 30
  [arXiv:1207.7235 [hep-ex]].


\bibitem{Higgs-fits}
T.~Corbett, O.~J.~P.~Eboli, J.~Gonzalez-Fraile and M.~C.~Gonzalez-Garcia,
  arXiv:1207.1344 [hep-ph];
P.~P.~Giardino, K.~Kannike, M.~Raidal and A.~Strumia,
  arXiv:1207.1347 [hep-ph];
J.~Ellis and T.~You,
  arXiv:1207.1693 [hep-ph];
M.~Montull and F.~Riva,
  arXiv:1207.1716 [hep-ph];
J.~R.~Espinosa, C.~Grojean, M.~Muhlleitner and M.~Trott,
  arXiv:1207.1717 [hep-ph];
D.~Carmi, A.~Falkowski, E.~Kuflik, T.~Volansky and J.~Zupan,
  arXiv:1207.1718 [hep-ph].



\bibitem{Arbey:2011ab}
  A.~Arbey, M.~Battaglia, A.~Djouadi, F.~Mahmoudi and J.~Quevillon,
  Phys.\ Lett.\ B {\bf 708} (2012) 162
  [arXiv:1112.3028 [hep-ph]].

\bibitem{SMSbounds}
  S.~Chatrchyan {\it et al.}  [CMS Collaboration],
  arXiv:1204.3774 [hep-ex];
  G.~Aad {\it et al.}  [ATLAS Collaboration],
  Phys.\ Lett.\ B {\bf 710} (2012) 67
  [arXiv:1109.6572 [hep-ex]];
  G.~Aad {\it et al.}  [ATLAS Collaboration],
  Phys.\  Rev.\  Lett.\  {\bf 108} (2012) 241802
  [arXiv:1203.5763 [hep-ex]];
  G.~Aad {\it et al.}  [ATLAS Collaboration],
  [arXiv:1203.6193 [hep-ex]];
G.~Aad {\it et al.}  [ATLAS Collaboration],
  Phys.\ Rev.\ Lett.\  {\bf 108} (2012) 181802
  [arXiv:1112.3832 [hep-ex]].

\bibitem{Chamseddine:1982jx}
  A.~H.~Chamseddine, R.~L.~Arnowitt and P.~Nath,
  Phys.\ Rev.\ Lett.\  {\bf 49} (1982) 970;
  R.~Barbieri, S.~Ferrara and C.~A.~Savoy,
  Phys.\ Lett.\ B {\bf 119} (1982) 343;
  L.~E.~Ibanez,
  Phys.\ Lett.\ B {\bf 118} (1982) 73;
  L.~J.~Hall, J.~D.~Lykken and S.~Weinberg,
  Phys.\ Rev.\ D {\bf 27} (1983) 2359;
  N.~Ohta,
  Prog.\ Theor.\ Phys.\  {\bf 70} (1983) 542.


\bibitem{Antusch:2011xz}
  S.~Antusch, L.~Calibbi, V.~Maurer, M.~Monaco and M.~Spinrath,
  arXiv:1111.6547 [hep-ph].

\bibitem{Abe:2007kf}
  H.~Abe, T.~Kobayashi and Y.~Omura,
  Phys.\ Rev.\  D {\bf 76} (2007) 015002
  [arXiv:hep-ph/0703044].

\bibitem{Gogoladze:2009bd}
  I.~Gogoladze, M.~U.~Rehman and Q.~Shafi,
  Phys.\ Rev.\ D {\bf 80} (2009) 105002
  [arXiv:0907.0728 [hep-ph]].


\bibitem{Horton:2009ed}
  D.~Horton, G.~G.~Ross,
  Nucl.\ Phys.\  {\bf B830 } (2010)  221-247.
  [arXiv:0908.0857 [hep-ph]].


\bibitem{Ghilencea:2012gz}
  D.~M.~Ghilencea, H.~M.~Lee and M.~Park,
  arXiv:1203.0569 [hep-ph].

\bibitem{Brummer:2012zc}
  F.~Brummer and W.~Buchmuller,
  JHEP {\bf 1205} (2012) 006
  [arXiv:1201.4338 [hep-ph]].

\bibitem{Hall:2011aa}
  L.~J.~Hall, D.~Pinner and J.~T.~Ruderman,
  JHEP {\bf 1204} (2012) 131
  [arXiv:1112.2703 [hep-ph]];
  A.~Arvanitaki and G.~Villadoro,
  JHEP {\bf 1202} (2012) 144
  [arXiv:1112.4835 [hep-ph]];
  Z.~Kang, J.~Li and T.~Li,
  arXiv:1201.5305 [hep-ph];
  M.~Asano and T.~Higaki,
  arXiv:1204.0508 [hep-ph];
  H.~M.~Lee, V.~Sanz and M.~Trott,
  JHEP {\bf 1205} (2012) 139
  [arXiv:1204.0802 [hep-ph]];
  J.~L.~Feng and D.~Sanford,
  arXiv:1205.2372 [hep-ph];
  K.~Blum, R.~T.~D'Agnolo and J.~Fan,
  arXiv:1206.5303 [hep-ph];
  M.~W.~Cahill-Rowley, J.~L.~Hewett, A.~Ismail and T.~G.~Rizzo,
  arXiv:1206.5800 [hep-ph];
  L.~Randall and M.~Reece,
  arXiv:1206.6540 [hep-ph];
  B.~Kyae and J.~-C.~Park,
  arXiv:1207.3126 [hep-ph];
  H.~Baer, V.~Barger, P.~Huang, A.~Mustafayev and X.~Tata,
  arXiv:1207.3343 [hep-ph];
  P.~Grothaus, M.~Lindner and Y.~Takanishi,
  arXiv:1207.4434 [hep-ph].



\bibitem{g-2}
  M.~Passera, W.~J.~Marciano, A.~Sirlin,
  Phys.\ Rev.\  {\bf D78 } (2008)  013009.
  [arXiv:0804.1142 [hep-ph]];
  K.~Hagiwara, R.~Liao, A.~D.~Martin, D.~Nomura, T.~Teubner,
  J.\ Phys.\ G {\bf G38}, 085003 (2011).
  [arXiv:1105.3149 [hep-ph]].

\bibitem{Barbieri:1987fn}
  R.~Barbieri, G.~F.~Giudice,
  Nucl.\ Phys.\  {\bf B306 } (1988)  63.

\bibitem{arXiv:1110.6926}
  M.~Papucci, J.~T.~Ruderman and A.~Weiler,
  arXiv:1110.6926 [hep-ph].

\bibitem{hep-ph/9303291}
  B.~de Carlos and J.~A.~Casas,
  Phys.\ Lett.\ B\ {\bf 309} (1993) 320
  [hep-ph/9303291];
  L.~Giusti, A.~Romanino and A.~Strumia,
  Nucl.\ Phys.\ B\ {\bf 550} (1999) 3
  [hep-ph/9811386];
  A.~Romanino and A.~Strumia,
  Phys.\ Lett.\ B\ {\bf 487} (2000) 165
  [hep-ph/9912301];
  J.~L.~Feng and K.~T.~Matchev,
  Phys.\ Rev.\ D {\bf 63} (2001) 095003
  [hep-ph/0011356].

\bibitem{PDG}
K.~Nakamura {\it et al.} [ Particle Data Group Collaboration ],
J.\ Phys.\ G {\bf G37}, 075021 (2010).


\bibitem{Djouadi:1998di}
  A.~Djouadi {\it et al.}  [MSSM Working Group Collaboration],
  hep-ph/9901246.

\bibitem{Martin:1993zk}
  S.~P.~Martin and M.~T.~Vaughn,
  Phys.\ Rev.\ D {\bf 50} (1994) 2282
   [Erratum-ibid.\ D {\bf 78} (2008) 039903]
  [hep-ph/9311340].

\bibitem{Aaij:2012ac}
  R.~Aaij {\it et al.}  [LHCb Collaboration],
  arXiv:1203.4493 [hep-ex].

\bibitem{Allanach:2001kg}
  B.~C.~Allanach,
  Comput.\ Phys.\ Commun.\  {\bf 143} (2002) 305
  [hep-ph/0104145].




\bibitem{focuspoint}
  K.~L.~Chan, U.~Chattopadhyay and P.~Nath,
  Phys.\ Rev.\ D {\bf 58} (1998) 096004
  [hep-ph/9710473];
  J.~L.~Feng, K.~T.~Matchev and T.~Moroi,
  Phys.\ Rev.\ Lett.\  {\bf 84} (2000) 2322
  [hep-ph/9908309];
  J.~L.~Feng, K.~T.~Matchev and T.~Moroi,
  Phys.\ Rev.\ D {\bf 61} (2000) 075005
  [hep-ph/9909334];
  S.~Akula, M.~Liu, P.~Nath and G.~Peim,
  Phys.\ Lett.\ B {\bf 709} (2012) 192
  [arXiv:1111.4589 [hep-ph]];
  S.~Akula, B.~Altunkaynak, D.~Feldman, P.~Nath and G.~Peim,
  Phys.\ Rev.\ D {\bf 85} (2012) 075001
  [arXiv:1112.3645 [hep-ph]].

\bibitem{Group:2010ab}
  T.~T.~E.~W.~Group [CDF and D0 Collaboration],
  arXiv:1007.3178 [hep-ex].




\bibitem{Georgi:1974sy}
  H.~Georgi and S.~L.~Glashow,
  Phys.\ Rev.\ Lett.\  {\bf 32} (1974) 438.

\bibitem{Pati:1974yy}
  J.~C.~Pati and A.~Salam,
  Phys.\ Rev.\ D {\bf 10} (1974) 275
   [Erratum-ibid.\ D {\bf 11} (1975) 703].

\bibitem{Georgi:1974my}
  H.~Georgi,
  AIP Conf.\ Proc.\  {\bf 23} (1975) 575;
  H.~Fritzsch and P.~Minkowski,
  Annals Phys.\  {\bf 93} (1975) 193.

\bibitem{Arbey:2011un}
 A.~Arbey, M.~Battaglia and F.~Mahmoudi,
 Eur.\ Phys.\ J.\ C {\bf 72} (2012) 1847
 [arXiv:1110.3726 [hep-ph]];
H.~K.~Dreiner, M.~Kr\"amer and J.~Tattersall,
 arXiv:1207.1613 [hep-ph].


\bibitem{Antusch:2011sq}
  S.~Antusch, L.~Calibbi, V.~Maurer, M.~Spinrath,
  Nucl.\ Phys.\  {\bf B852 } (2011)  108-148.
  [arXiv:1104.3040 [hep-ph]].

\bibitem{arXiv:0902.4644}
  S.~Antusch and M.~Spinrath,
  Phys.\ Rev.\ D\ {\bf 79} (2009) 095004
  [arXiv:0902.4644 [hep-ph]].

\bibitem{SUSYthresholds}
 L.~J.~Hall, R.~Rattazzi and U.~Sarid,
 Phys.\ Rev.\  D {\bf 50} (1994) 7048
 [arXiv:hep-ph/9306309];
 M.~S.~Carena, M.~Olechowski, S.~Pokorski and C.~E.~M.~Wagner,
 Nucl.\ Phys.\  B {\bf 426} (1994) 269
 [arXiv:hep-ph/9402253];
 R.~Hempfling,
 Phys.\ Rev.\  D {\bf 49} (1994) 6168;
 T.~Blazek, S.~Raby and S.~Pokorski,
 Phys.\ Rev.\  D {\bf 52} (1995) 4151
 [arXiv:hep-ph/9504364].

\bibitem{Freitas:2007dp}
  A.~Freitas, E.~Gasser and U.~Haisch,
  Phys.\ Rev.\  D {\bf 76} (2007) 014016
  [arXiv:hep-ph/0702267].

\bibitem{Antusch:2008tf}
  S.~Antusch, M.~Spinrath,
  Phys.\ Rev.\  {\bf D78 } (2008)  075020.
  [arXiv:0804.0717 [hep-ph]].

\bibitem{Spinrath:2010dh}
  M.~Spinrath,
  arXiv:1009.2511 [hep-ph].

\bibitem{Cassel:2009ps}
  S.~Cassel, D.~M.~Ghilencea, G.~G.~Ross,
  Nucl.\ Phys.\  {\bf B825 } (2010)  203-221.
  [arXiv:0903.1115 [hep-ph]];
  S.~Cassel, D.~M.~Ghilencea, G.~G.~Ross,
  Phys.\ Lett.\  {\bf B687 } (2010)  214-218.
  [arXiv:0911.1134 [hep-ph]];
  S.~Cassel, D.~M.~Ghilencea and G.~G.~Ross,
  Nucl.\ Phys.\  B {\bf 835} (2010) 110
  [arXiv:1001.3884 [hep-ph]];
  S.~Cassel, D.~M.~Ghilencea, S.~Kraml, A.~Lessa, G.~G.~Ross,
  JHEP {\bf 1105 } (2011)  120.
  [arXiv:1101.4664 [hep-ph]].

\bibitem{SuperIso}
 F.~Mahmoudi,
 Comput.\ Phys.\ Commun.\  {\bf 178} (2008) 745
 [arXiv:0710.2067 [hep-ph]];
 F.~Mahmoudi,
 Comput.\ Phys.\ Commun.\  {\bf 180} (2009) 1579
 [arXiv:0808.3144 [hep-ph]].

\bibitem{Asner:2010qj}
  D.~Asner {\it et al.}  [Heavy Flavor Averaging Group Collaboration],
  arXiv:1010.1589 [hep-ex].

\bibitem{Barberio:2008fa}
  E.~Barberio {\it et al.}  [Heavy Flavor Averaging Group Collaboration],
  arXiv:0808.1297 [hep-ex].


\bibitem{g-2exp_jparc}
  T.~Mibe [J-PARC g-2 Collaboration],
  Chin.\ Phys.\ C {\bf 34} (2010) 745.

\bibitem{g-2exp_fermilab}
  R.~M.~Carey, K.~R.~Lynch, J.~P.~Miller, B.~L.~Roberts, W.~M.~Morse, Y.~K.~Semertzides, V.~P.~Druzhinin and B.~I.~Khazin {\it et al.},
  FERMILAB-PROPOSAL-0989.

\bibitem{Lees:2012ju}
  J.~P.~Lees {\it et al.}  [BABAR Collaboration],
  arXiv:1207.0698 [hep-ex].

\bibitem{SMS}
  J.~Alwall, P.~Schuster and N.~Toro,
  Phys.\ Rev.\ D {\bf 79} (2009) 075020
  [arXiv:0810.3921 [hep-ph]].
  D.~S.~M.~Alves, E.~Izaguirre and J.~G.~Wacker,
  JHEP {\bf 1110} (2011) 012
  [arXiv:1102.5338 [hep-ph]].
  D.~Alves {\it et al.}  [LHC New Physics Working Group Collaboration],
  arXiv:1105.2838 [hep-ph].


\bibitem{gauginoratios}
  J.~Chakrabortty and A.~Raychaudhuri,
  Phys.\ Lett.\ B {\bf 673} (2009) 57
  [arXiv:0812.2783 [hep-ph]];
  S.~P.~Martin,
  Phys.\ Rev.\ D {\bf 79} (2009) 095019
  [arXiv:0903.3568 [hep-ph]].

  
\bibitem{Arbey:2012dq}
  A.~Arbey, M.~Battaglia, A.~Djouadi and F.~Mahmoudi,
  arXiv:1207.1348 [hep-ph].
  
\bibitem{Brignole:1993dj}
  A.~Brignole, L.~E.~Ibanez and C.~Munoz,
  Nucl.\ Phys.\ B {\bf 422} (1994) 125
   [Erratum-ibid.\ B {\bf 436} (1995) 747]
  [hep-ph/9308271].
  
\end{thebibliography}
\end{document}